\documentclass[a4paper,11pt]{article}
\usepackage[latin1]{inputenc}
\usepackage[english]{babel}
\usepackage{txfonts}
\usepackage{latexsym}
\usepackage{graphicx}
\usepackage[square]{natbib} 
\bibpunct{(}{)}{;}{a}{}{,}
\begin{document}
\title{Branes: cosmological surprise and observational deception}
\author{Stéphane Fay\\
School of Mathematical Science,\\
Queen Mary University of London, Mile End road, London E1 4NS, UK\\
and\\
Laboratoire Univers et Théories (LUTH), UMR 8102\\
Observatoire de Paris, F-92195 Meudon Cedex, France\\
\small{steph.fay@gmail.fr}}
\maketitle
\abstract{
Using some supernovae and CMB data, we constrain the Cardassian, Randall-Sundrum, and Dvali-Gabadadze-Porrati brane-inspired cosmological models. We show that a transient acceleration and an early loitering period are usually excluded by the data. Moreover, the three models are equivalent to some usual quintessence/ghost dark energy models defined by a barotropic index $\gamma_\phi$ depending on the redshift. We calculate this index for each model and show that they mimic a universe close to a $\Lambda CDM$ model today.
}
\section{Introduction}\label{s0}
The notion of extradimensions (see \cite{AbeMar00} for a popularised review) takes root in the search for a unified theory of fundamental interactions. In the twenties, \cite{Kal21} tried to unify electromagnetism with Einstein's general relativity (GR) thanks to a geometrical scheme involving a compactified extradimension. However, such a solution does not work with weak and strong interactions. Interest in extradimensions returned in the seventies and the eighties with the development of string theory and supersymmetry, here most symmetric forms occur for a ten or eleven dimensional universe. In particular, in the mid eighties, Michael Green and John Schwartz discovered the superstring theory that is able to incorporate quantum mechanics without any problem of infinity, provided we live in a ten-dimensional universe, that is, our four-dimensional universe with six compactified extradimensions. 

For all these theories, it is assumed that the four fundamental interactions are everywhere in the ten-dimensional spacetime. A breakthrough came in the late nineties when one wonder if gravitational interaction could be the only one able to travel in the ten dimensions while the other interactions would stay confined in our four-dimensional brane world. One of the interesting things about the brane idea is that it could explain why gravity is so weak with respect to the other interactions. For instance, it could be diluted in the extradimensions as explained by \cite{ArkDimDva98}. Another interest of branes could be to explain why the cosmological constant $\Lambda$ is so small. Randall and Sundrum looked for branes in which the bulk dimensions are strongly curved or warped but not necessarily compactified. Then, if they are warped in the right way, it is possible to explain the smallness of $\Lambda$.

As a result, branes offer many new and exciting possibilities. Some of them, related to the universe expansion, have been qualified "cosmological surprise" by \cite{Sah05}. For instance, some brane theories are able to produce a so-called transient acceleration, i.e. an acceleration of the expansion that is not eternal. Such behaviour would allow some string theories problems related to the S-Matrix to be avoided. Another cosmological surprise is the existence of some early loitering periods as defined by \cite{SahSht05}, that is some early periods of time during which the Hubble function grows more slowly than the Hubble function of the $\Lambda CDM$ model. Thus \cite{SahSht05} mathematically define the loitering as the existence of a minimum value of the ratio $H(z)/H_{\Lambda CDM}$. As a result, the age of a loitering universe is more than of a $\Lambda CDM$ model. It also increases the growth rate of density inhomogeneities with respect to the $\Lambda CDM$ model. Hence, an early epoch of loitering should help to boost the formation of high-redshift, gravitationally bound systems as noted by \cite{Sah05}.\\
The aim of this paper is to constrain the free parameters of three brane theories, the so-called Cardassian, Randall-Sundrum, and Dvali-Gabadadze-Porrati models (DGP). We use some supernovae and CMB data. We show that a transient acceleration and an early loitering period are excluded by the observations most of the time. Moreover, we relate each of these models to some of the usual quintessence or ghost dark energy theories; so we begin by introducing each of the three brane theories.\\
The Cardassian model was introduced for the first time by \cite{ChuFre00}. The authors consider a $3+1$ brane located at the $Z_2$ symmetry fixed plane of a $Z_2$ symmetric five-dimensional spacetime. With a suitable (and non unique) choice of the bulk energy momentum tensor, they found that on the brane, the Hubble function is proportional to a power of the matter density. The cosmological consequences of the expansion acceleration were analysed by \cite{FreLew02}. A fluid interpretation of the Cardassian model was also developed by \cite{GonFre03} as a description of the matter with some interaction terms which would give birth to an effective negative pressure able to drive the expansion acceleration.\\
There are two types of Randall-Sundrum models\citep{RanSun99A}. In the Randall-Sundrum type $I$ model\citep{Gum03}, two $3+1$ branes are embedded in a five-dimensional anti-de-Sitter ($AdS_5$) bulk with a negative cosmological constant. The extra dimension has $Z_2$ symmetry of about $y=L$ (where the brane stands with a negative tension where we live) and $y=0$ (where the other brane stands with a positive tension). We are interested by the Randall-Sundrum type $II$ model where the brane in $y=L$ is sent to infinity, whereas our brane is in $y=0$. It has a positive tension; otherwise, the Newton constant is negative.\\
Last, the Dvali-Gabadadze-Porrati (DGP) model\citep{DvaGabPor00} rests upon a $3+1$ brane in a five-dimensional Minkowski space with an infinite-size extra dimension. There is neither a bulk cosmological constant nor any brane tension.\\
The plan of the paper is the following. In Sect. \ref{s1}, we define the geometry and the matter content of our brane universe. In Sect. \ref{s2}-\ref{s4}, we use some observational data to constrain the Cardassian, Randall-Sundrum, and Dvali-Gabadadze-Porrati brane models. We conclude in Sect. \ref{s5}.
\section{Geometrical and physical frameworks}\label{s1}
\begin{figure}[h]
\centering
\includegraphics[width=12cm]{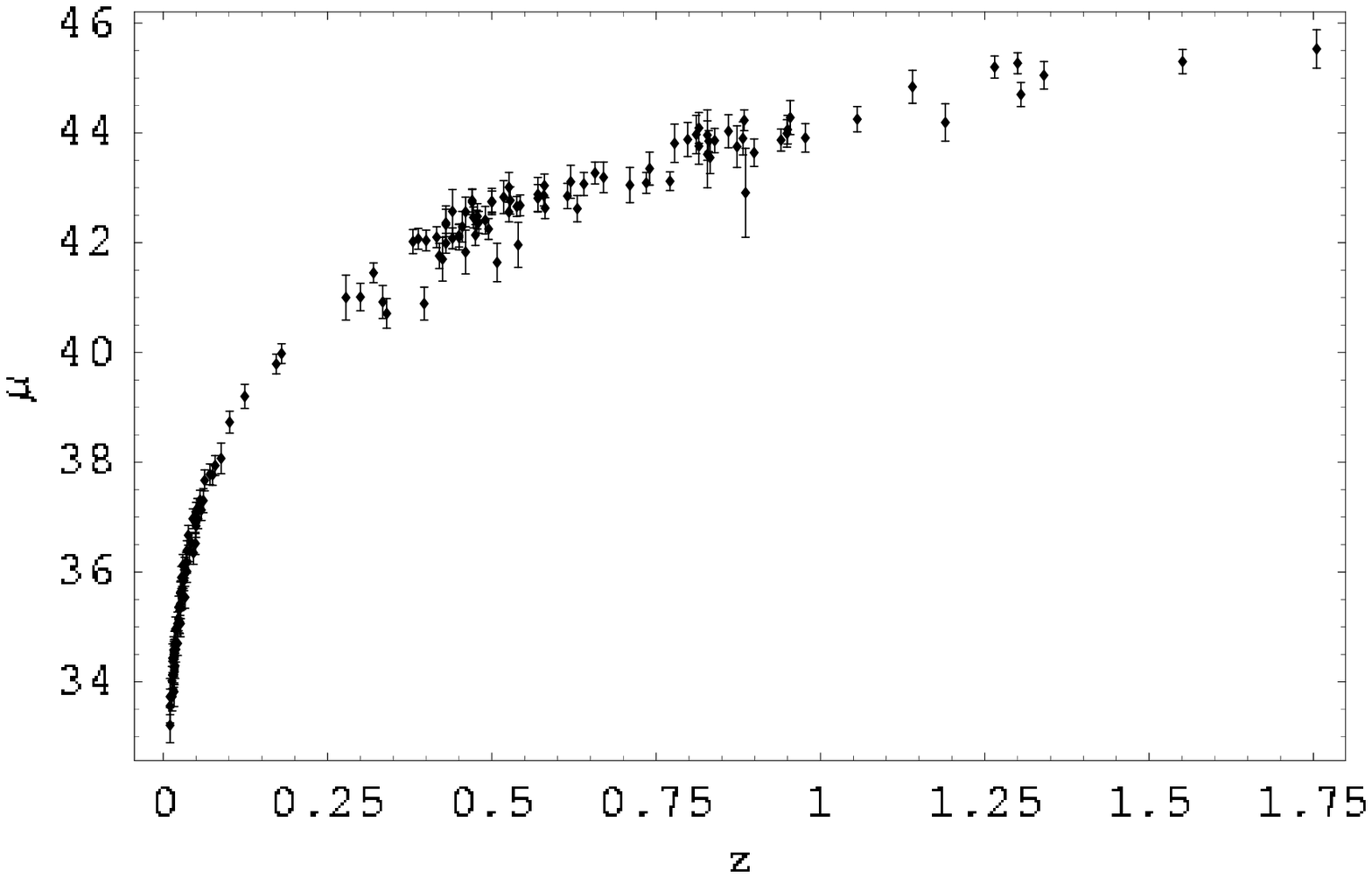}
\caption{\scriptsize{\label{dataRiess04}Magnitude versus redshift of the 157 supernovae of the Riess gold sample}}
\end{figure}

As a geometrical framework, we assume an isotropic and homogeneous universe described by a Friedmann-Lemaître-Robertson-Walker (FLRW) metric:
$$
ds^2=-dt^2+a(t)^2(\frac{dr^2}{1-kr^2}+r^2(d\theta^2+sin^2\theta d\phi^2))
$$
where $a(t)$ is the scale factor accounting for the universe's expansion or contraction. The parameter $k$ is $0$, $-1$, or $1$ if the universe geometry is Euclidean, hyperbolic or spherical respectively.\\
As a physical framework, we consider the Cardassian, Randall-Sundrum type $II$, and DGP brane models. On the brane, we assume the presence of a perfect fluid of matter with an equation of state (eos)
$$
p_m=(\gamma-1)\rho_m
$$
and a perfect fluid of dark energy with a constant eos
$$
p_e=(\Gamma-1)\rho_e.
$$
We choose $\gamma=1$, that is, a dust fluid (cold dark matter (CDM)). Dark energy will be a quintessence if $\Gamma>0$, and otherwise ghost.\\
Each model is defined by some parameters, such as the energy density parameter $\Omega_{m_0}$ of the CDM(see definition below). In what follows we look for the values of these parameters, allowing 
\begin{itemize}
\item an accelerated expansion for a redshift $z_a$. This redshift is such that the second derivative of the metric function $a$ with respect to the proper time $t$, $\ddot a$, vanishes.
\item a transient acceleration. This is when the acceleration does not last forever but only a finite time. It occurs when several values of $z$ are solutions to $\ddot a=0$.
\item a loitering period. Following \cite{SahSht05} (see their appendix), a minimum value of the ratio $H(z)/H_{\Lambda CDM}$ is a generic feature of a loitering universe. The loiteirng redshift $z_l$ is thus defined as the time of this minimum, such as $d(H(z)/H_{\Lambda CDM})/dz=0$. Note that other definitions of a loitering period exist, such that $dH(z)/dz=0$. Sometimes a loitering universe is also compared (although different) to the hesitating universe of Lemaître which is such that the second derivative of the scale factor vanishes in $z_l$.
\end{itemize}
In some cases, we also look for the domination redshift $z_d$ defined as the redshift, for which the linear term of the CDM, $\rho_m\propto (1+z)^3$, stops dominating the dynamics of the universe. We constrain the branes theories using the $157$ supernovae of the Riess gold sample\citep{Rie04}. It is plotted in Fig. \ref{dataRiess04}. To reach this goal, we use two $\chi^2$ tests with and without marginalising the Hubble constant. Since the supernovae data alone are not enough to avoid some large degeneracy, we also use some priors coming from the CMB data, more precisely, the WMAP prior $\Omega_{m_0}=0.27\pm 0.05$ got by considering dark energy defined by a constant eos. Is it appropriate? Most of the models of the present paper tend to a universe dominated by CDM at early-time like the model used in the WMAP analysis. Hence, we will naively suppose that the domination of the CDM at early-times should imply a similar prior to the one above. Moreover, it is widely used in the literature and will allow us to compare our results with the ones in other papers, even if this is not perfect. A more satisfactory analysis would require modifying a code like CMBFast to implement the brane models and then get some more realistic CMB priors. However, such a work is beyond the scope of this paper.\\
All the statistical procedures are described in appendix \ref{l1}. In the whole paper, we choose the positive value of the square root unless otherwise stated.
\section{Cardassian theory}\label{s2}
\subsection{Field equations}
The field equations of the Cardassian theory write
\begin{equation}\label{car1}
H^2+\frac{k}{a^2}=k_1(\rho_m+\rho_e)\left[1+A(\rho_m+\rho_e)^\nu\right]
\end{equation}
\begin{equation}\label{car2}
\dot H-\frac{k}{a^2}=-\frac{3}{2}k_1(\gamma\rho_m+\Gamma\rho_e)\left[1+A(\nu+1)(\rho_m+\rho_e)^\nu\right]
\end{equation}
with $\nu$ a constant and $k_1=\frac{8\pi G}{3}$. The conservation of the energy momentum tensor for the CDM and dark energy leads to
$$
\rho_m=\rho_{m_1}a^{-3\gamma}
$$
$$
\rho_e=\rho_{e_1}a^{-3\Gamma}
$$
with for the present time (indicated by the subscript 0) $\rho_{m_0}=\rho_{m_1}a_0^{-3\gamma}$ and $\rho_{e_0}=\rho_{e_1}a_0^{-3\Gamma}$. We define
\begin{itemize}
\item the redshift $1+z=\frac{a_0}{a}$
\item the CDM density parameter $\Omega_{m_0}=\frac{k_1\rho_{m_0}}{H_0^2}$
\item the dark energy density parameter $\Omega_{e_0}=\frac{k_1\rho_{e_0}}{H_0^2}$
\item the curvature density parameter $\Omega_{k_0}=-\frac{k}{a_0^2H_0^2}$
\item and last $\Omega_{\nu_0}=\frac{k_1^\nu}{AH_0^{2\nu}}$
\end{itemize}
Equations (\ref{car1}) and (\ref{car1}+\ref{car2}) write then as
\begin{eqnarray}\label{car3}
H^2&=&H_0^2\left[\Omega_{m_0}(1+z)^{3\gamma}+\Omega_{e_0}(1+z)^{3\Gamma}\right]\nonumber\\
&&\left[1+\frac{1}{\Omega_{\nu_0}}\left[\Omega_{m_0}(1+z)^{3\gamma}+\Omega_{e_0}(1+z)^{3\Gamma}\right]^\nu\right]+\nonumber\\
&&H_0^2\Omega_{k_0}(1+z)^2
\end{eqnarray}
\begin{eqnarray}\label{car4}
\frac{1}{H_0^2}\frac{\ddot a}{a}&=&\Omega_{m_0}(1-\frac{3}{2}\gamma)(1+z)^{3\gamma}+\Omega_{e_0}(1-\frac{3}{2}\Gamma)(1+z)^{3\Gamma}+\nonumber\\
&&\frac{1}{\Omega_{\nu_0}}\left[\Omega_{m_0}(1+z)^{3\gamma}+\Omega_{e_0}(1+z)^{3\Gamma}\right]^\nu\nonumber\\
&&\mbox{[}\Omega_{m_0}(1-\frac{3}{2}(\nu+1)\gamma)(1+z)^{3\gamma}+\nonumber\\
&&\Omega_{e_0}(1-\frac{3}{2}(\nu+1)\Gamma)(1+z)^{3\Gamma}\mbox{]}
\end{eqnarray}
with the constraint
\begin{equation}\label{cardCons}
\Omega_{\nu_0}=\frac{(\Omega_{m_0}+\Omega_{e_0})^{\nu+1}}{1-\Omega_{m_0}-\Omega_{e_0}-\Omega_{k_0}}
\end{equation}
where $H_0$ is the present value of the Hubble constant. In \cite{FayFuzAli05}, we constrained six models of quintessence/ghost dark energy. All of them are equivalent to the presence of a minimally coupled and massive scalar field $\phi$ in the universe. Here we show that such models may also mimic the Cardassian model of this section (and the following, see below!). Following the calculations of \cite{FayFuzAli05}, it is possible to show that the above Cardassian model is equivalent to the usual GR in four dimensions with quintessence/ghost dark energy whose eos $p_\phi=(\gamma_\phi-1)\rho_\phi$ would be defined by the varying barotropic index
\begin{eqnarray*}
\gamma_\phi&=&\frac{(1+\nu)F(z)^\nu\left[\Omega_{e_0}\Gamma(1+z)^{3\Gamma}+\Omega_{m_0}\gamma(1+z)^{3\gamma}\right]}{F(z)^{1+\nu}+\Omega_{e_0}\Omega_{\nu_0}(1+z)^{3\Gamma}}+\\
&&\frac{\Omega_{e_0}\Omega_{\nu_0}\Gamma(1+z)^{3\Gamma}}{F(z)^{1+\nu}+\Omega_{e_0}\Omega_{\nu_0}(1+z)^{3\Gamma}}\\
\end{eqnarray*}
with $F(z)=\Omega_{e_0}(1+z)^{3\Gamma}+\Omega_{m_0}(1+z)^{3\gamma}$. The present value of dark energy density parameter would then be defined by $\Omega_{\phi_0}=\Omega_{e_0}+\frac{(\Omega_{e_0}+\Omega_{m_0})^{1+\nu}}{\Omega_{\nu_0}}$.
\subsection{Flat universe and CDM}
\begin{figure}[h]
\centering
\includegraphics[width=4cm]{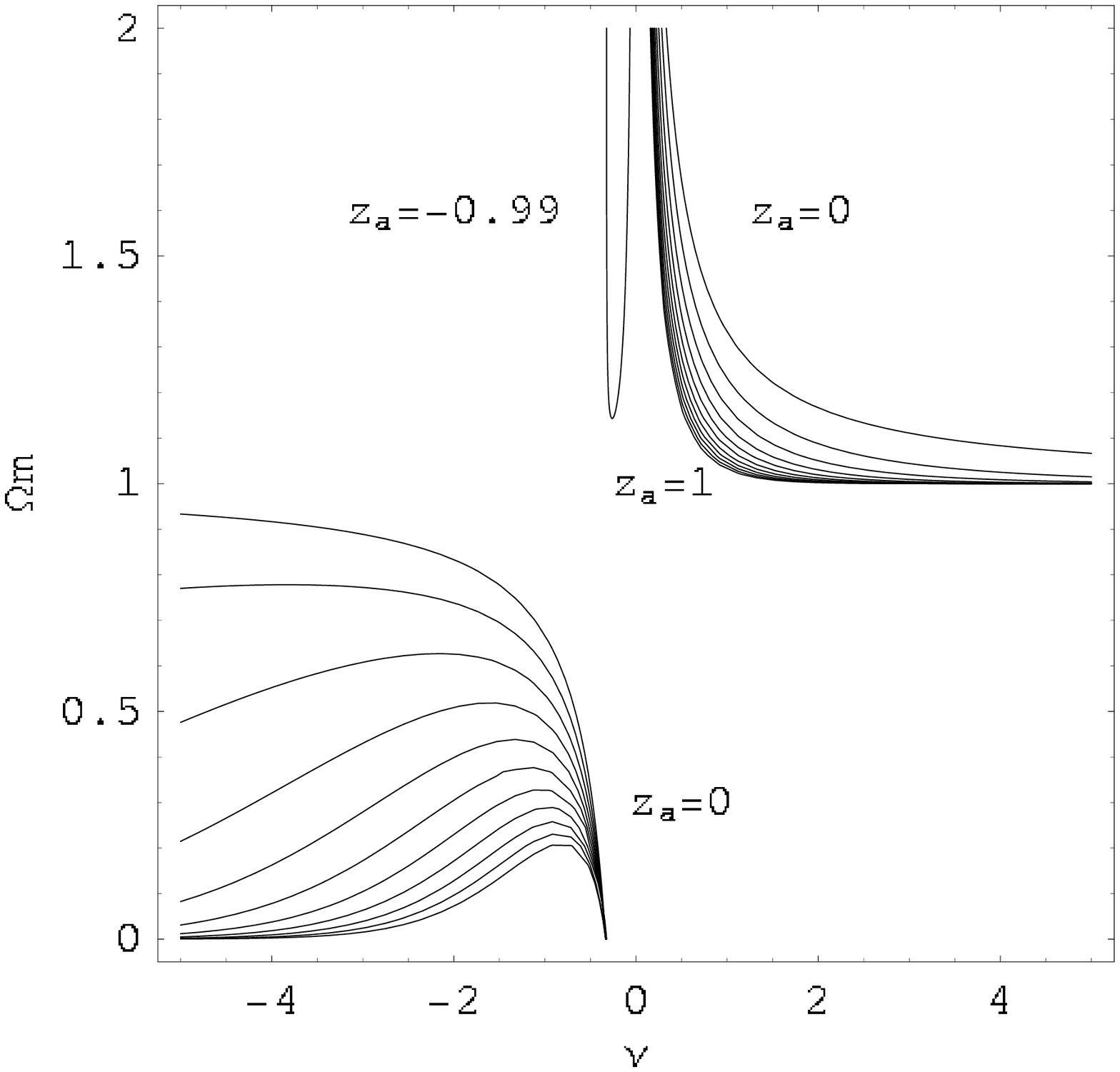}
\includegraphics[width=4cm]{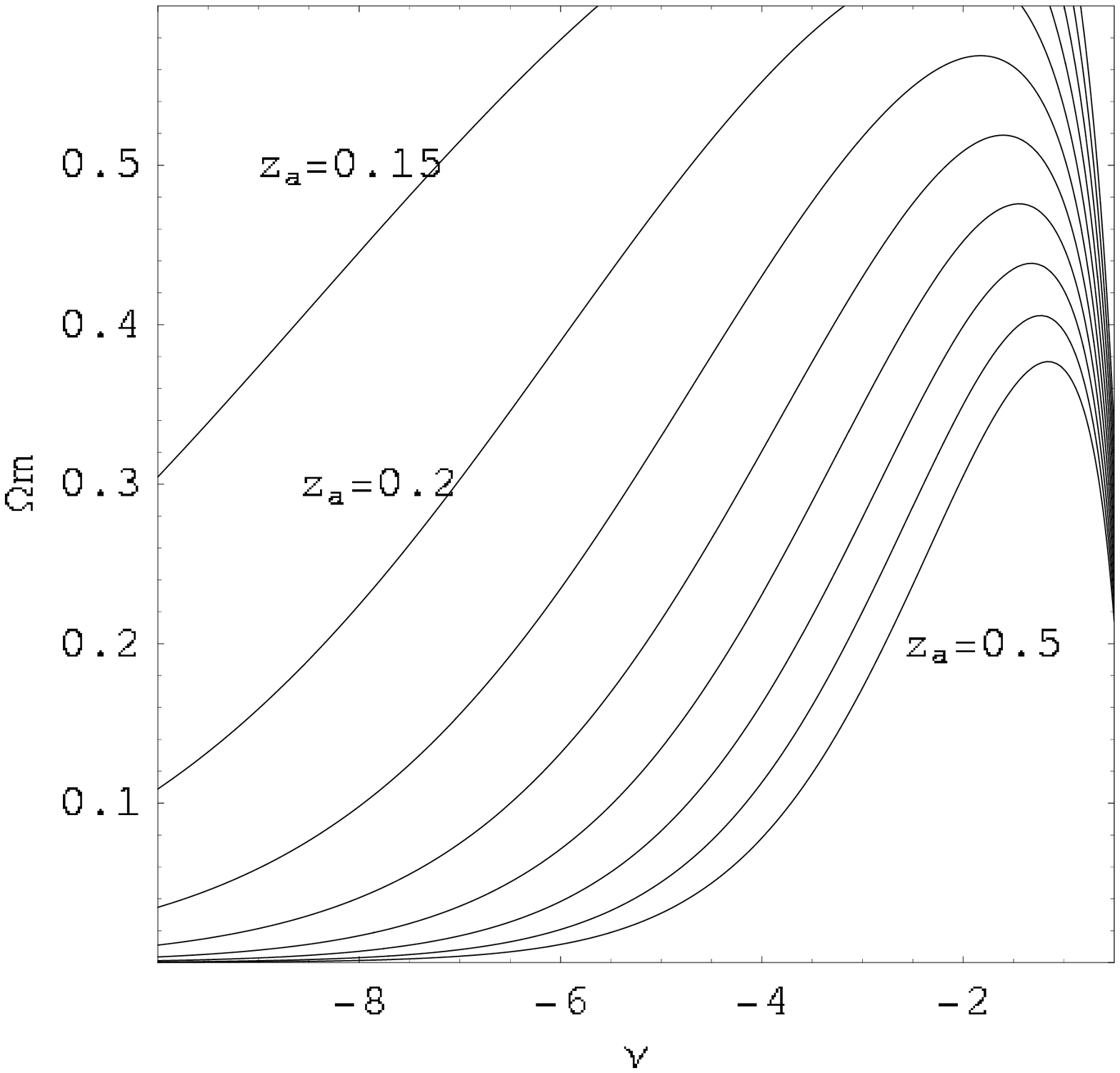}
\includegraphics[width=4cm]{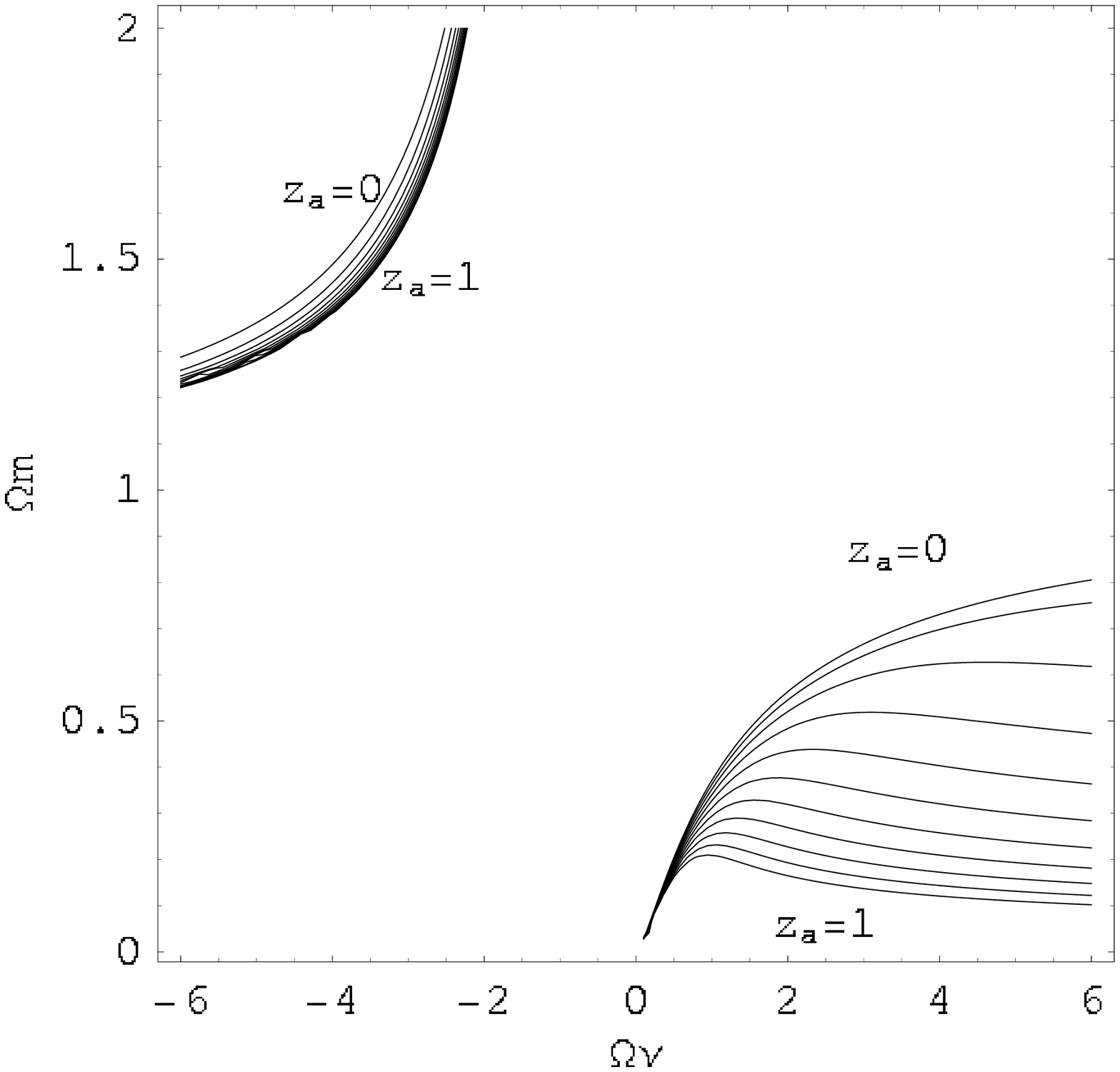}
\caption{\scriptsize{\label{cardFlatza}The acceleration redshift for the Cardassian theory with CDM and a flat universe in the $(\nu,\Omega_{m_0})$ and $(\Omega_{\nu_0},\Omega_{m_0})$. The second graph is an enhancement of the first one in the region of space parameters covered by the $2\sigma$ confidence contour obtained with the supernovae data.}}
\end{figure}

\begin{figure}[h]
\centering
\includegraphics[width=4.3cm]{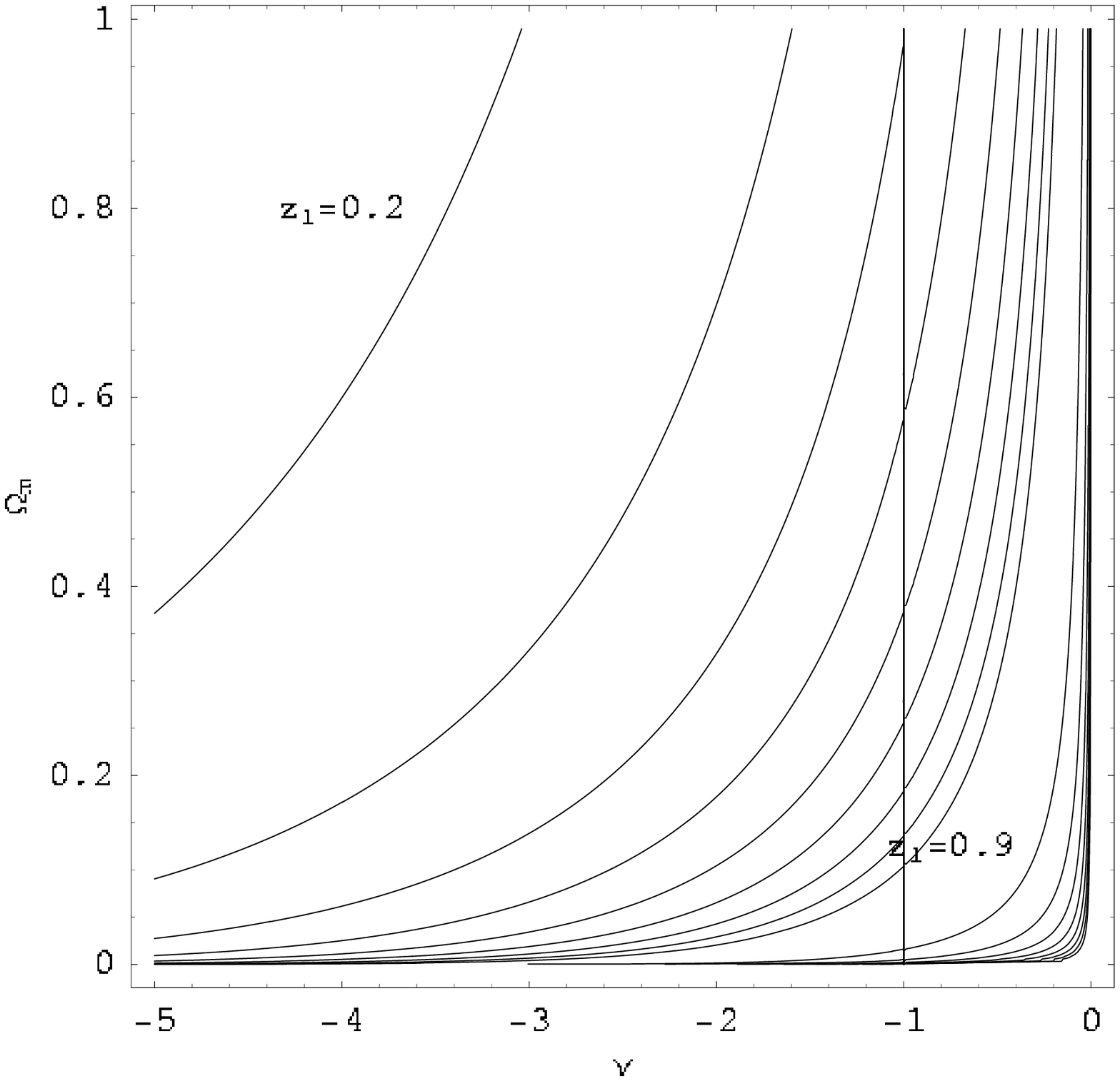}
\caption{\scriptsize{\label{cardFlatzl}Some loitering redshift from $z_l=0.2$ to $z_l$ equal to several unities for the Cardassian theory with CDM in a flat universe.}}
\end{figure}

\begin{figure}[h]
\centering
\includegraphics[width=4cm]{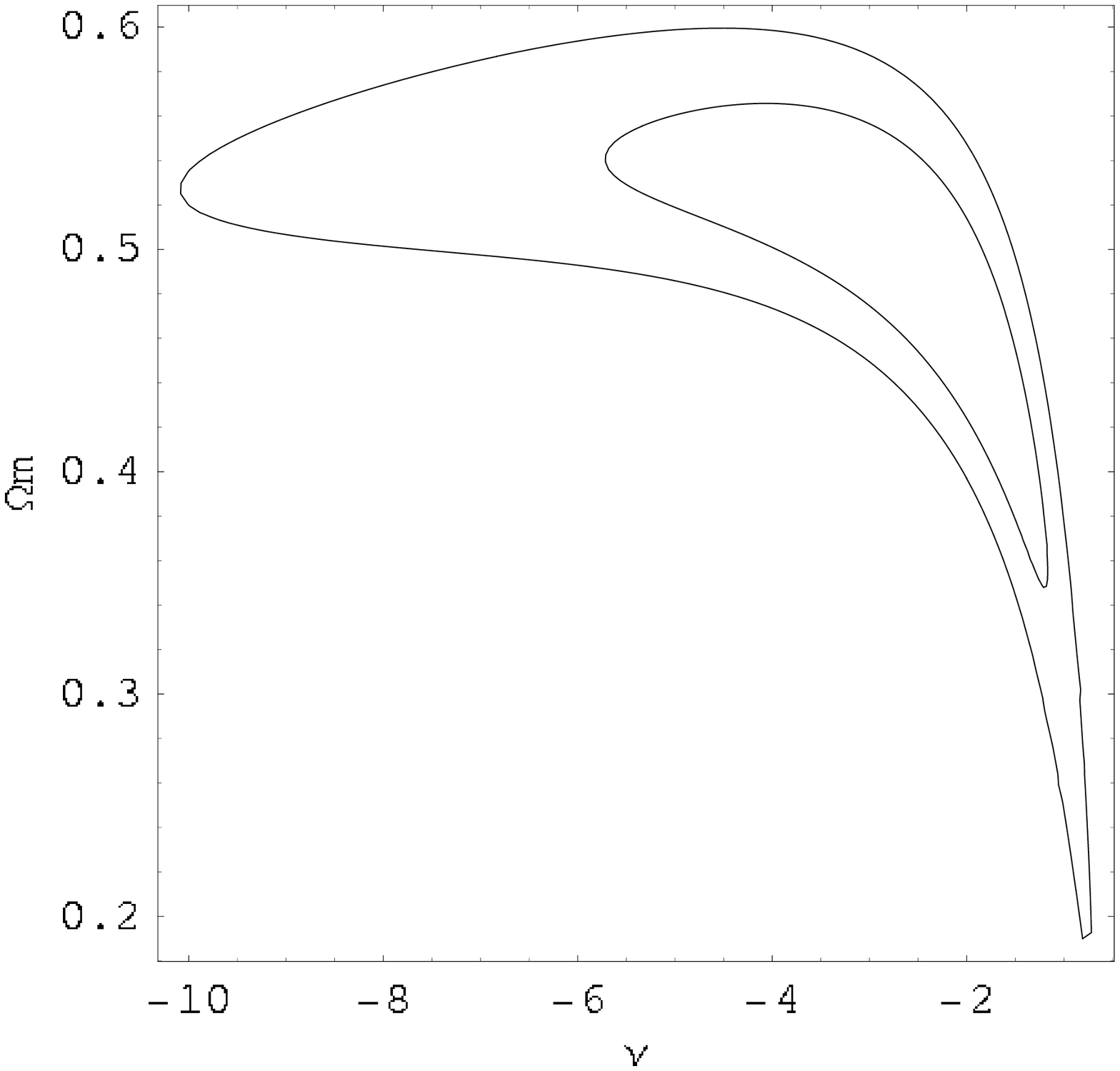}
\includegraphics[width=4cm]{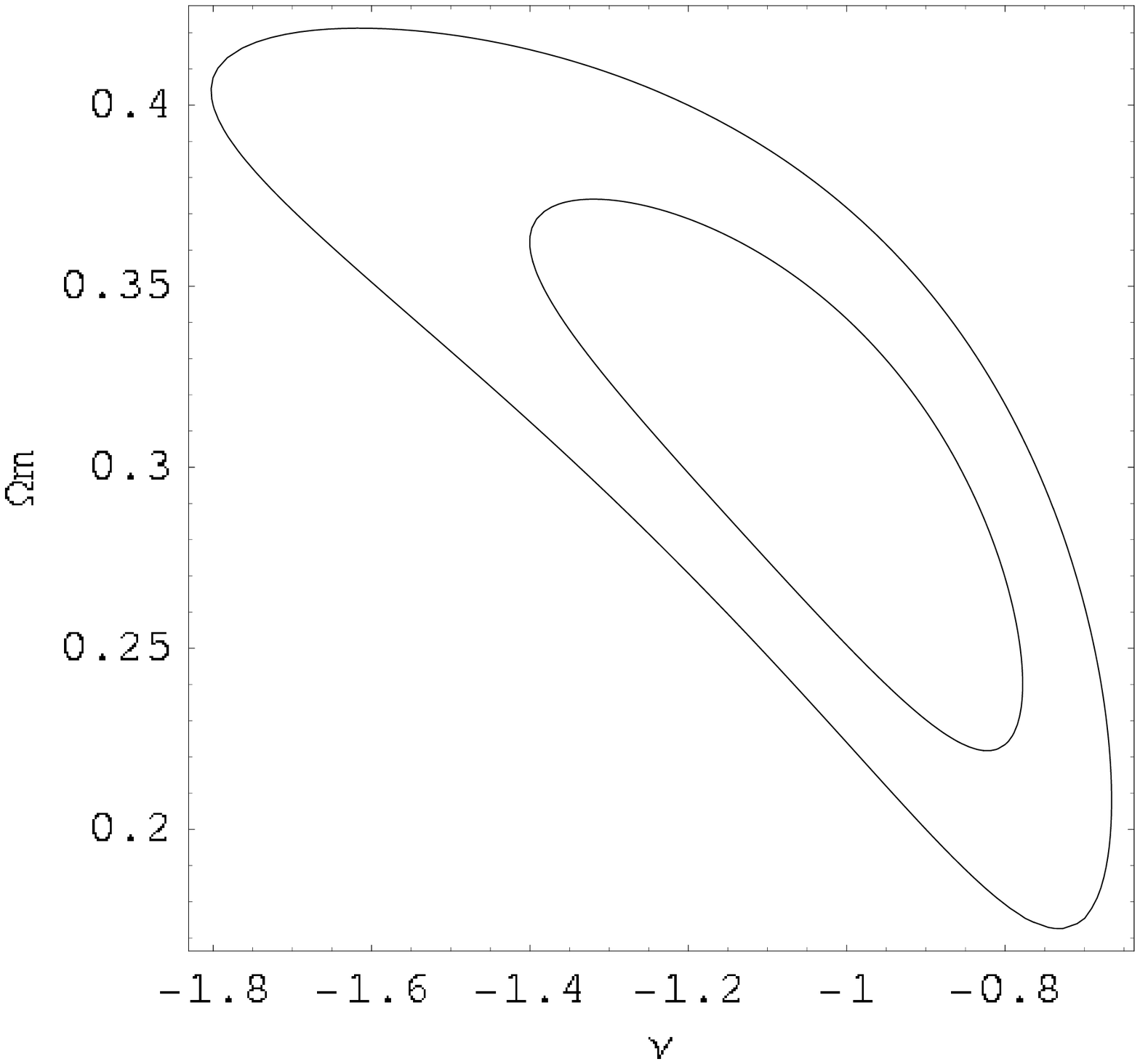}
\includegraphics[width=4cm]{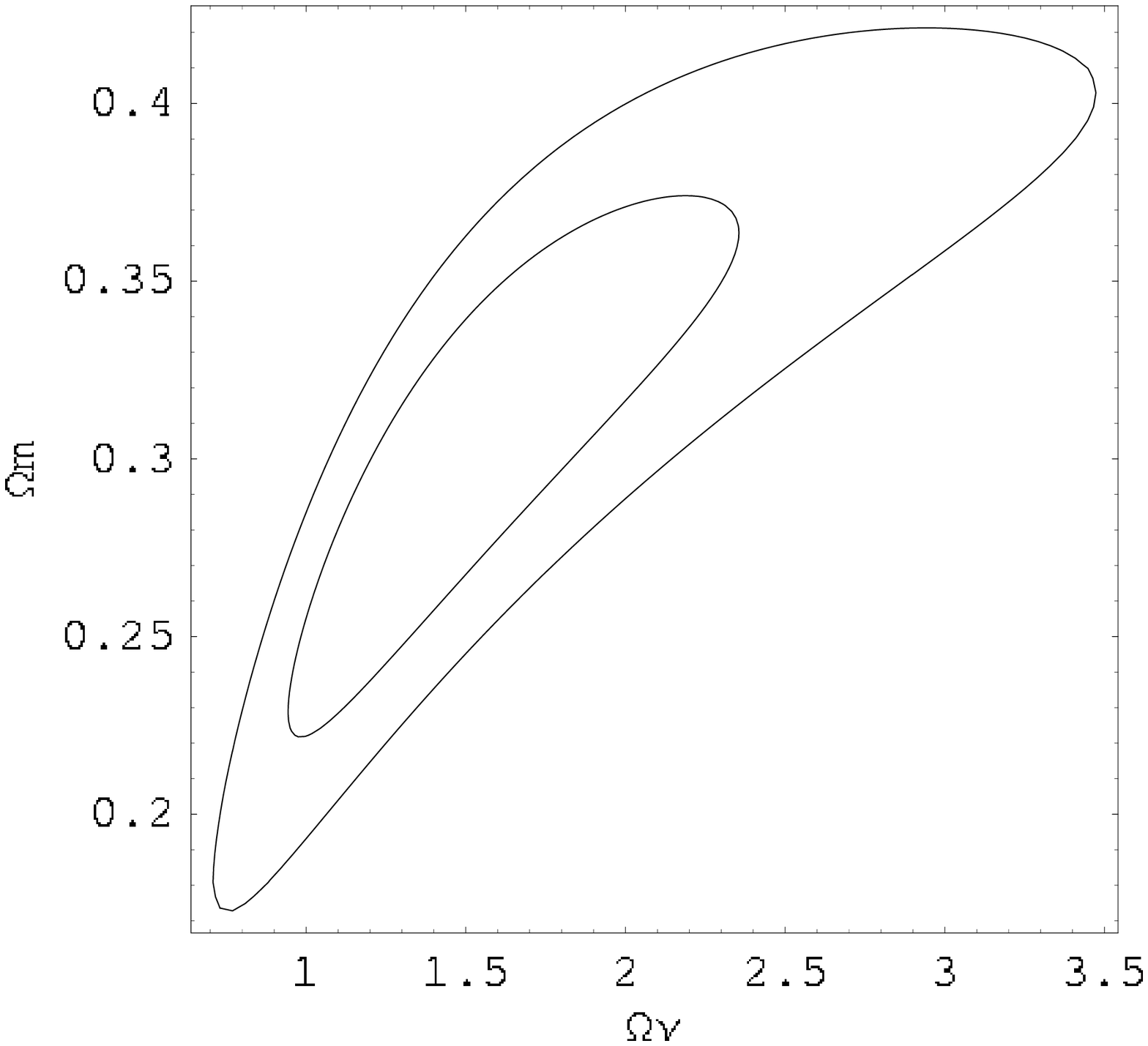}
\caption{\scriptsize{\label{cardFlat}$1\sigma$ and $2\sigma$ confidence contours for the flat Cardassian model with CDM. The first graph is obtained with supernovae data. The two last one take into account supernovae and CMB data by assuming the prior $\Omega_{m_0}=0.27\pm 0.05$.}}
\end{figure}

If there is no dark energy on the brane ($\Omega_{e_0}=0$) but only some CDM, it becomes $\gamma_\phi=\gamma(1+\nu)$ whatever the $\Omega_{k_0}$. The Cardassian theory is thus equivalent to GR in four dimensions with dark energy with a constant eos\footnote{This is true from the supernovae viewpoint but not from the perturbation viewpoint.}. In this case, the acceleration of the expansion arises if $\nu<2/(3\gamma)-1$, a Big-Rip may occur if $\nu<-1$, and the model is equivalent to a $\Lambda CDM$ model if $\nu=-1$. The $\gamma_\phi=const$ case has already been studied in several papers. What is interesting here is the value of $\Omega_{\nu_0}$ as determined by the observations.\\
We begin by studying the theory mathematically. In the presence of curvature, the acceleration redshift $z_a$ of the Cardassian theory reads
\begin{equation}\label{car5}
z_a=-1+\Omega_{m_0}^{-\frac{1}{3\gamma}}\left[\frac{\Omega_{\nu_0}(2-3\gamma)}{3\gamma(1+\nu)-2}\right]^{\frac{1}{3\gamma\nu}}
\end{equation}
This expression depends on the curvature via $\Omega_{\nu_0}$ and the expression (\ref{cardCons}). Clearly when $\gamma=1$, the values $\nu=0$ (CDM universe), $\nu=-1/3$, and $\Omega_{\nu_0}$ play a special role as shown below.\\
The solutions of Eq. \ref{car5}) for the flat universe ($\Omega_{k_0}=0$) considered in this subsection are plotted in Fig. \ref{cardFlatza}. As shown by Eq. \ref{car5}), there are four groups of solutions (putting $\gamma=1$):
\begin{itemize}
\item one with $\Omega_{\nu_0}<0$ ($\Omega_{m_0}>1$) and $-1/3<\nu<0$ ($z_a$ is real): the universe expansion first decelerates and then accelerates in the future (see the curves labeled by $z_a=-0.99$ on the first graph of Fig. \ref{cardFlatza}). The constant $A$ or the parameter $\Omega_{\nu_0}$ are negative. The late time acceleration is due to the domination of the negative Cardassian term (equivalent to the presence of dark energy with $\gamma_\phi$) in the future ($z_a$ close to $-1$) in Eq. \ref{car4}).
\item one with $\Omega_{\nu_0}<0$ ($\Omega_{m_0}>1$) and $\nu>0$ ($z_a$ is real): the universe expansion first accelerates and then decelerates. The constant $A$ or the parameter $\Omega_{\nu_0}$ is negative. The early-time acceleration is due to the domination of the negative Cardassian term (equivalent to the presence of dark energy with $\gamma_\phi$) at early-times in Eq. \ref{car4}).
\item one with $\Omega_{\nu_0}>0$ ($\Omega_{m_0}<1$) and $\nu<-1/3$ ($z_a$ is real): the universe expansion first decelerates and then accelerates. The constant $A$ or the parameter $\Omega_{\nu_0}$ is positive. The late time acceleration is due to the domination of the negative Cardassian term (equivalent to the presence of dark energy with $\gamma_\phi$) at late times in Eq. \ref{car4}).
\item one with $-\Omega_{\nu_0}$ and $\nu+1/3$ having opposite signs ($z_a$ is complex). Then there is no transition between a decelerated and an accelerated expansion.
\end{itemize}
Since the expansion is accelerating, this is the third group of solutions that interests us in the framework of this paper. Since none of the curves in Fig. \ref{cardFlatza} crosses each other, $\ddot a$ never vanishes twice for the same values of the constant parameters $(\Omega_{m_0},\nu)$ and thus the acceleration is not a transient phenomenon.\\
The loitering redshift as a function of $\Omega_{m_0}$ and $\nu$ is plotted in Fig. \ref{cardFlatzl}: early-time loitering (some few units) occurs for low values of $\Omega_{m_0}$ whatever $\nu$ or low values of $\nu$ whatever $\Omega_{m_0}$.
\\
What do the supernovae data tell us about this Cardassian model? When we marginalise the Hubble constant, the smallest $\chi^2$ is $176.39$ when $\Omega_{m_0}=0.49$ and $\nu=-2.40$. The $\chi^2$ per degrees of freedom is $\chi^2_{DOF}=1.13$, which is reasonable. We find that $\Omega_{\nu_0}=5.32$ and that the brane mimics the presence of ghost dark energy defined by $\gamma_\phi<0$. At $2\sigma$, we find $\Omega_{m_0}\in\left[0.2,0.6\right]$ and $\nu\in\left[-10.1,-0.75\right]$. The $1\sigma$ and $2\sigma$ confidence contours are plotted on the left graph of Fig. \ref{cardFlat}. Unhappily, the $\Omega_{\nu_0}$ parameter is very degenerate since it may be as large as $750$ at $2\sigma$.\\
Minimalising $\chi^2$ with $H_0$, we find that its lowest value occurs for $H_0=65.6$\footnote{We do not give the $2\sigma$ contours of $H_0$ since most of time, we have more than three free parameters and the degeneracy is too large to be meaningful.} (with the same values for the other free parameters $\Omega_{m_0}$ and $\nu$ that have been determined by marginalisation - see Appendix \ref{l1} for an explanation). Then, the age of the universe is $13.26$ billion years and the acceleration of the expansion takes place when $z_a=0.29$.
\\
We try to better constrain the $\Omega_{\nu_0}$ parameter by assuming the WMAP prior $\Omega_{m_0}=0.27\pm 0.05$. It could be justified by the fact that the linear term of the CDM, $\rho_m\propto(1+z)^3$, dominates in the field equations at early-times. Then, the smallest value for $\chi^2$ is $\chi^2=180.14$ ($\chi^2_{DOF}=1.16$, still reasonable!) with $\Omega_{m_0}=0.29$ and $\nu=-1.01$. It corresponds to $\Omega_{\nu_0}=1.42$ and a Cardassian model mimicking ghost dark energy very close to a $\Lambda CDM$ model $(\nu=-1)$. At $2\sigma$, $\Omega_{m_0}\in\left[0.17,0.42\right]$ and $\nu\in\left[-1.8,-0.7\right]$. The $2\sigma$ confidence contour of the $\Omega_{\nu_0}$ parameter is now physically meaningful since $\Omega_{\nu_0}\in\left[0.7,3.47\right]$. The confidence contours are plotted on the two last graphs of the Fig. \ref{cardFlat}. Frith, using the Doppler peaks in the CMB, $230$ supernovae collated from the literature and some assumptions about upper limits at which the Cardassian term begins to dominate the expansion, \cite{Fri04} finds $\Omega_{m_0}\in\left[0.19,0.26\right]$ and $\nu\in\left[-0.99, -0.76\right]$. Minimalisation gives $H_0=64.6$ for the Hubble constant and thus the age of the universe would be $14.64$ billion years. The acceleration of the expansion begins at $z_a=0.69$. This is a higher value than with the supernovae data alone. The cold dark matter begins to dominate around $z_d=0.34$ and thus, the prior we assume on $\Omega_{m_0}$ is justified.\\
We see here how a precise determination of $\Omega_{m_0}$ is important. If one does not consider the WMAP results for this parameter, the supernovae data alone predict a value of $\Omega_{m_0}$ that is higher than $0.27$, and $\Omega_{\nu_0}$ is degenerated. Then, the Cardassian model mimics ghost dark energy. If one takes the WMAP results $\Omega_{m_0}=0.27$ into account, the Cardassian model is very similar to a $\Lambda CDM$ model and $\Omega_{\nu_0}$ is better constrained. This shows how a precise measurement of $\Omega_{m_0}$ could help us to determine what makes the expansion accelerate. A loitering period is compatible with the data only at late times, for $z_l<1$.
\subsection{Curved universe and CDM}\label{s23}
\begin{figure*}[h]
\centering
\includegraphics[width=12cm]{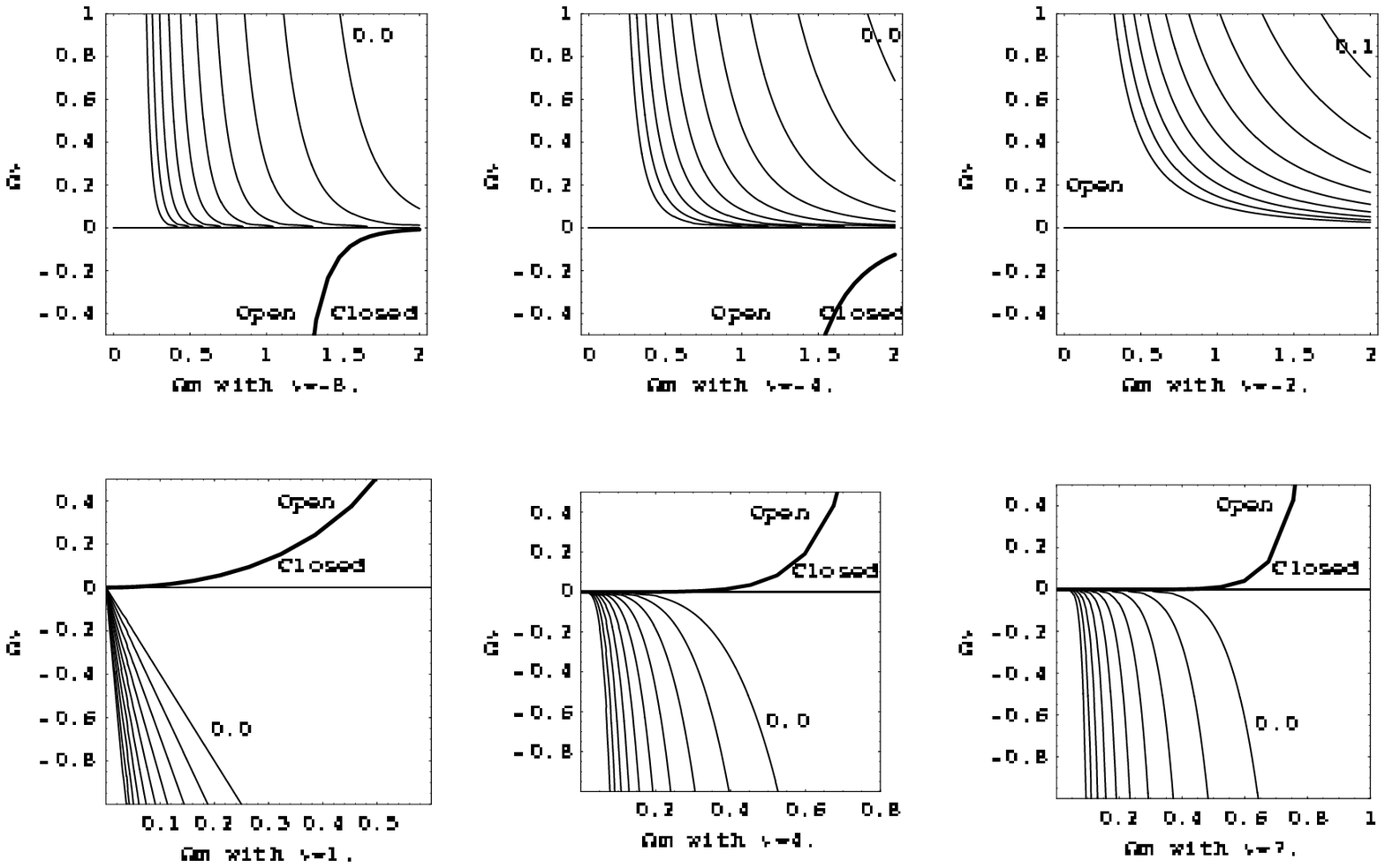}
\caption{\scriptsize{\label{cardza}The acceleration redshift for the Cardassian theory with CDM in a curved universe. The curves on each graph correspond (labels from the right to the left) to $z_a=0.0, 0.1, ..., 1.0$.}}
\end{figure*}

\begin{figure}[h]
\centering
\includegraphics[width=6cm]{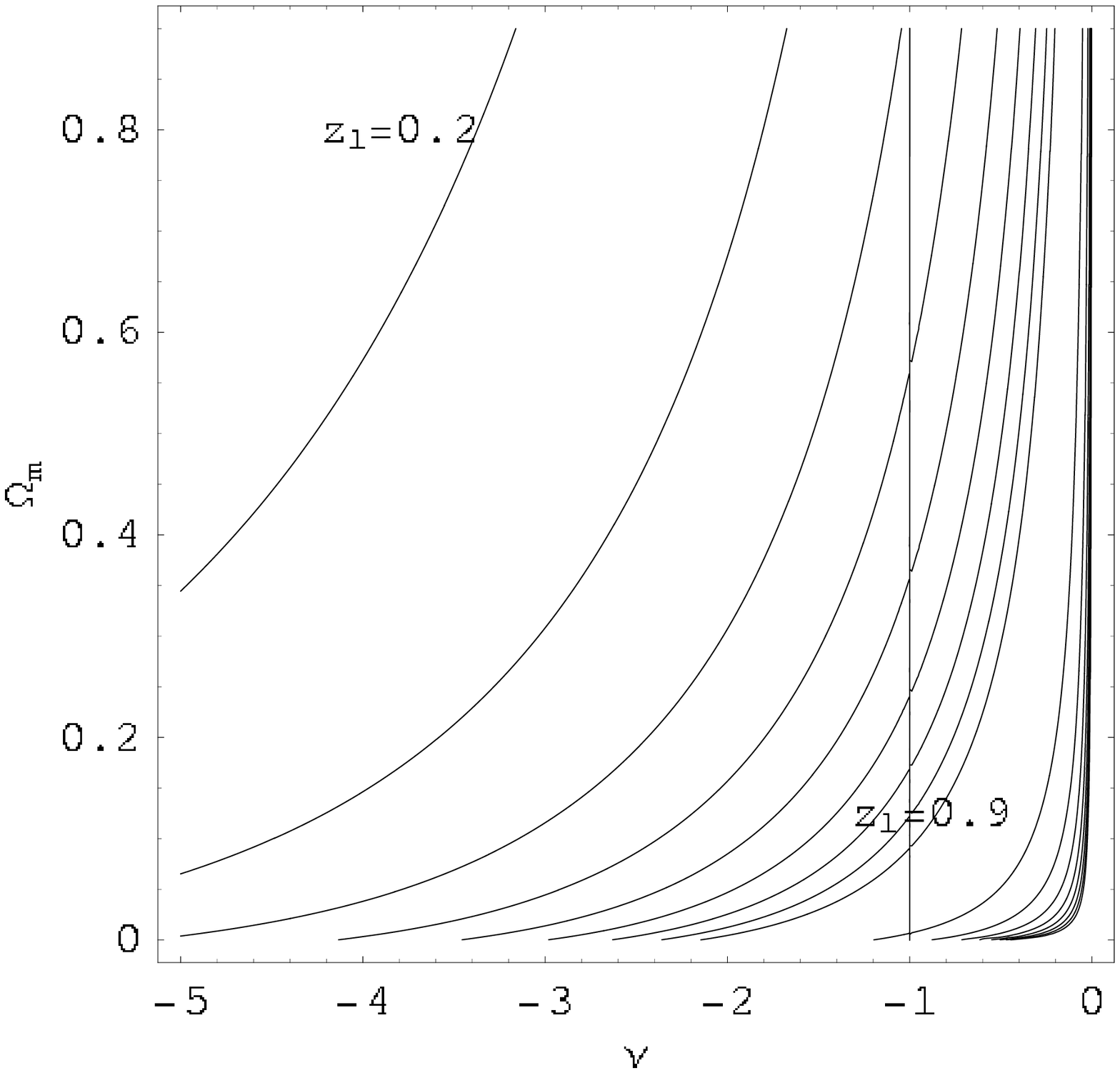}
\caption{\scriptsize{\label{cardzl}The loitering redshift for the Cardassian theory with CDM in a curved universe ($\Omega_{k_0}=0.05$).}}
\end{figure}

\begin{figure*}[h]
\centering
\includegraphics[width=12cm]{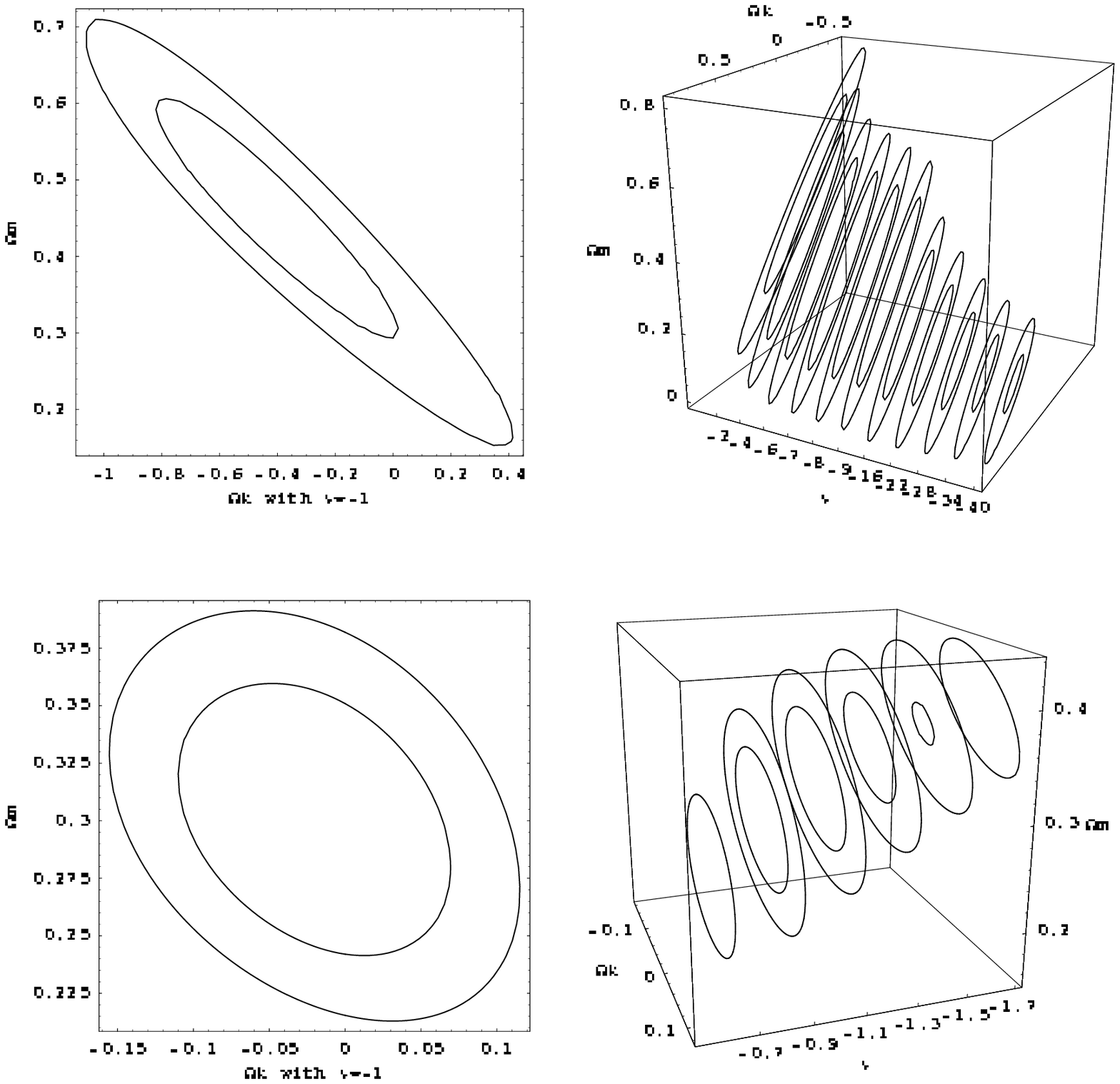}
\caption{\scriptsize{\label{cardCur}$1\sigma$ and $2\sigma$ confidence contours (for $\nu=-1$) and volumes for the Cardassian model with CDM in a curved universe. The first two graphs are without prior on $\Omega_{m_0}$ and $\Omega_{k_0}$. The last two graphs are such that $\Omega_{m_0}=0.27\pm 0.05$ and $\Omega_{k_0}=-0.02\pm 0.05$.}}
\end{figure*}

The solutions (\ref{car5}) for $z_a$ in the presence of curvature are plotted in the Fig. \ref{cardza}. The remarks of the previous subsection about the solutions (\ref{car5}) stay valid with $\Omega_{m_0}+\Omega_{k_0}$ instead of $\Omega_{m_0}$ alone. In the reasonably large region of space parameters that we have explored, the acceleration is never transient: no curve crosses an other in Fig. \ref{cardza}.\\
The solutions for the loitering redshift are similar to the one plotted in Fig. \ref{cardFlatzl}. But the higher is $\Omega_{k_0}$ (i.e, ranging from negative to positive values), the lower $\nu$ nedd to be to allow loitering as illustrated in Fig. \ref{cardzl}.
\\Using the supernovae data alone, the smallest $\chi^2$ after marginalisation on the Hubble constant is $\chi^2=176.4$ ($\chi^2_{DOF}=1.14$) with $\Omega_{m_0}=0.49$, $\Omega_{k_0}=0.0$, and $\nu=-2.38$. Hence it predicts a flat universe\footnote{Indeed, if we want to stay realistic, finding a perfectly flat universe with some imperfect data probably means that the universe is curved...} and the results for the values of the $\Omega_{\nu_0}$ parameter, the acceleration redshift $z_a$, the Hubble constant, and the universe age are the same as in the previous section. The $1\sigma$ and $2\sigma$ confidence contours are plotted in the upper left graph of Fig. \ref{cardCur}. At $2\sigma$, one has $\nu\in\left[-0.7,-70\right]$, $\Omega_{m_0}\in\left[0.1,0.75\right]$ and $\Omega_{k_0}\in\left[-1.2,0.7\right]$. 
\\Once again, at early-times, the linear term of the CDM density $(1+z)^3$ dominates and thus seems a reasonable prior for assuming that $\Omega_{m_0}=0.27\pm 0.05$. The best fit now corresponds to $\Omega_{m_0}=0.27$ but with $\Omega_{k_0}=0.40$ and $\nu=-9.05$ and $\Omega_{k_0}$ is thus too large to be realistic! Hence, we assume an additional prior inspired from WMAP results, i.e. a slightly closed universe $\Omega_{k_0}=-0.02\pm 0.05$. In this case we get $\chi^2=180.12$ ($\chi^2_{DOF}=1.17$) with $\Omega_{m_0}=0.30$, $\Omega_{k_0}=-0.02$ and $\nu=-0.98$. Then, we deduce $\Omega_{\nu_0}=1.35$. The Cardassian model mimics quintessence dark energy, once again close to a $\Lambda CDM$ model. $1\sigma$ and $2\sigma$ confidence contours are plotted on the right graphs in Fig. \ref{cardCur}. At $2\sigma$, one has $\nu\in\left[-2.13,-0.63\right]$, $\Omega_{m_0}\in\left[0.16,0.43\right]$ and $\Omega_{k_0}\in\left[-0.16,0.12\right]$. Minimalising, we find that $H_0=64.6$, that is, $14.63$ billion years old for the universe. A late ($z_l<1$) loitering period agrees with the supernovae data but almost excluded if we also take the CMB data into account. Acceleration occurs when $z_a=0.69$. The CDM begins to dominate when $z_d=0.32$, thus justifying the WMAP prior on $\Omega_{m_0}$.\\
Note that the Cardassian model in a flat and curved universe has also been studied in \cite{God04} with a different statistical procedure and with other supernovae data than the ones used in the present paper. Some similar results have been got with some similar priors.
\subsection{Flat universe and dark energy on the brane}
\begin{figure*}[h]
\centering
\includegraphics[width=12cm]{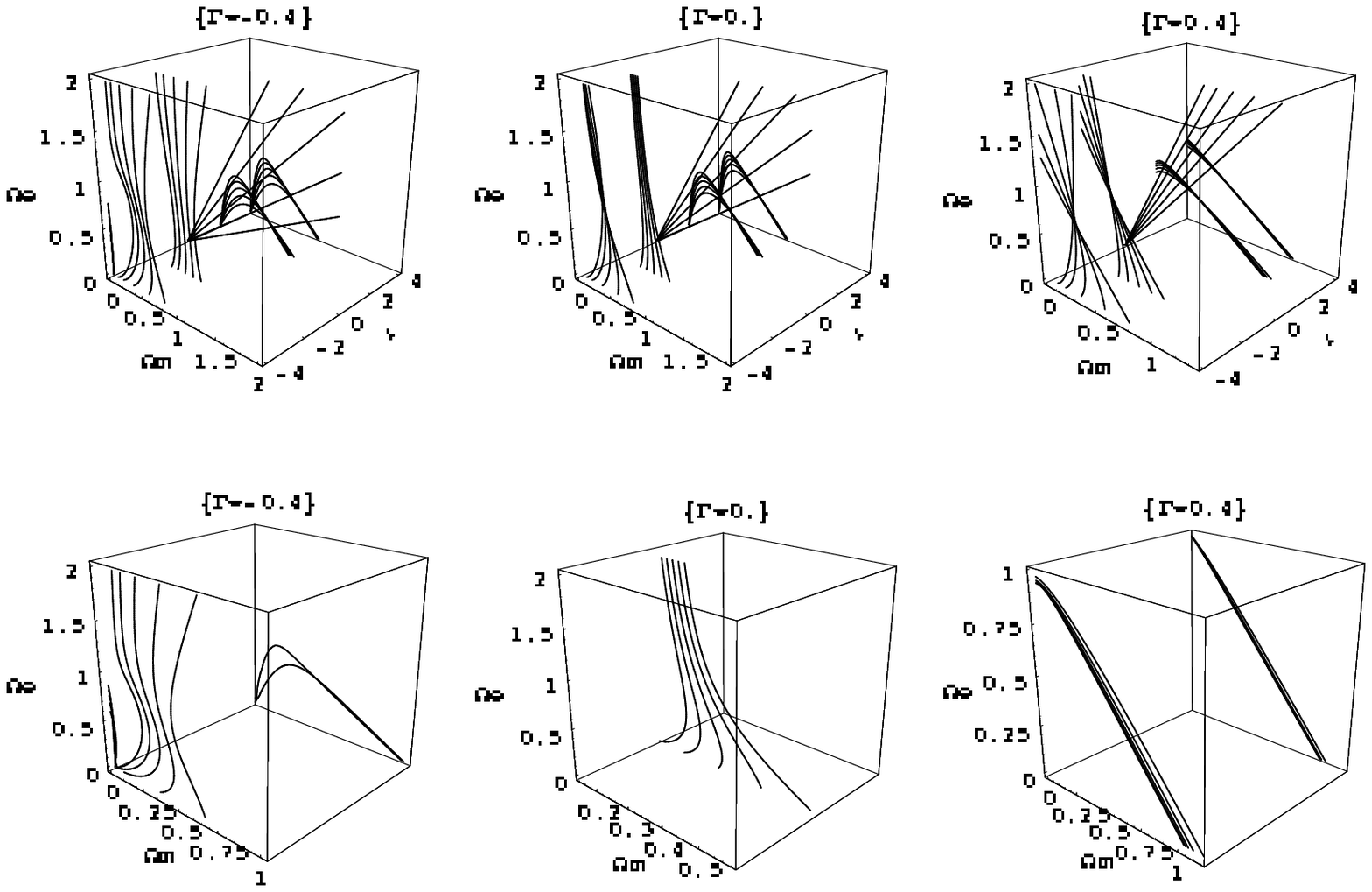}
\caption{\scriptsize{\label{cardFlatDEza}The acceleration redshift for the Cardassian theory with CDM and dark energy in a flat universe. The curves in the planes labelled by $\nu$ correspond to $z_a=0.1, 0.2 ... 0.5$. The last three graphs are the solutions plotted on the first three graphs and they correspond to a universe that is first decelerated and presently accelerating.}}
\end{figure*}

\begin{figure*}[h]
\centering
\includegraphics[width=12cm]{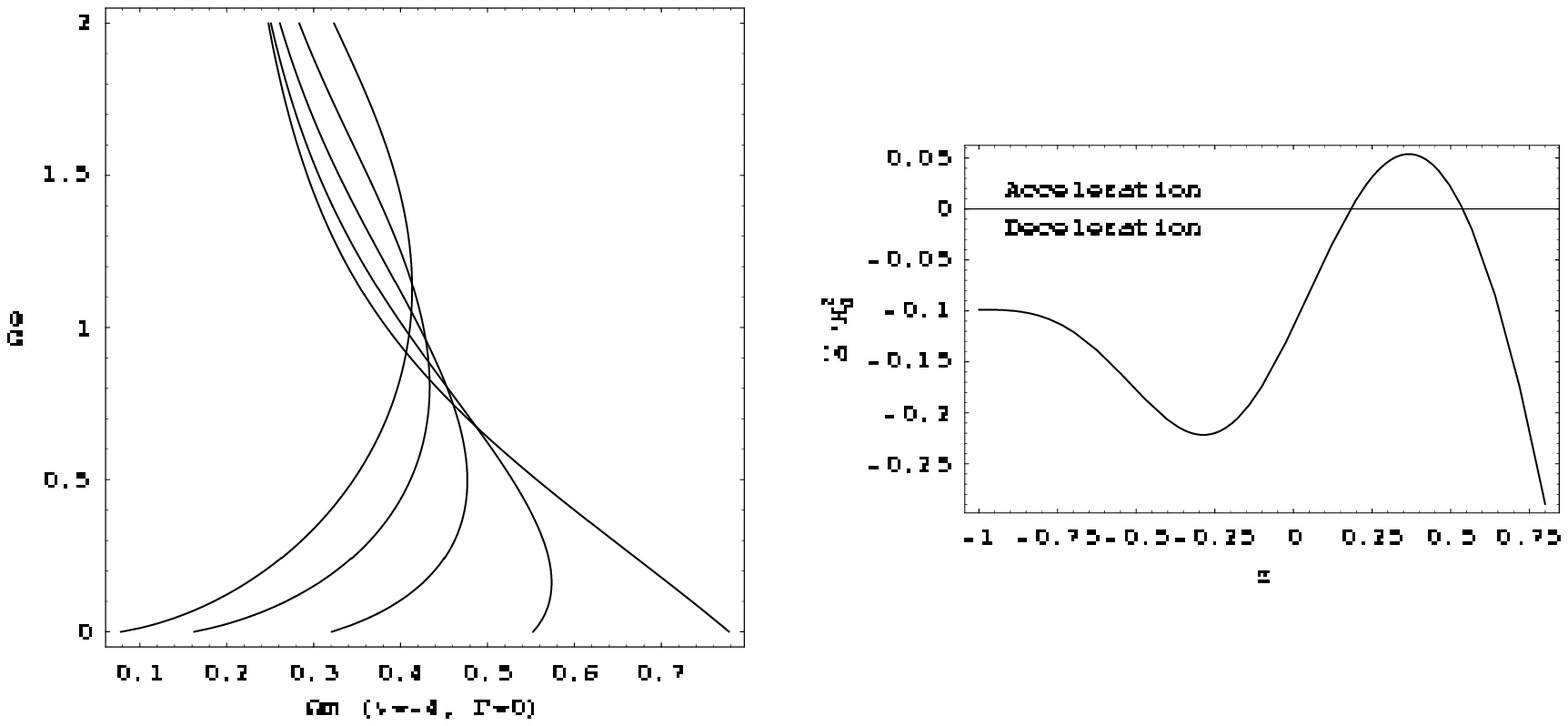}
\caption{\scriptsize{\label{cardFlatDETrans}Existence of transient acceleration for the Cardassian theory with CDM and dark energy in a flat universe. The first graph shows some solutions when $\Gamma=0$ and $\nu=-4$. Some curves, parameterised by a value of $z_a$, cross, showing the presence of some field equations solutions describing a transient acceleration (transient deceleration is also possible but not here). One of these solutions is plotted on the right graph when $(\Omega_{m_0},\Omega_{e_0},\Gamma,\nu)=(0.4,1,0,-4)$.}}
\end{figure*}

\begin{figure}[h]
\centering
\includegraphics[width=4.3cm]{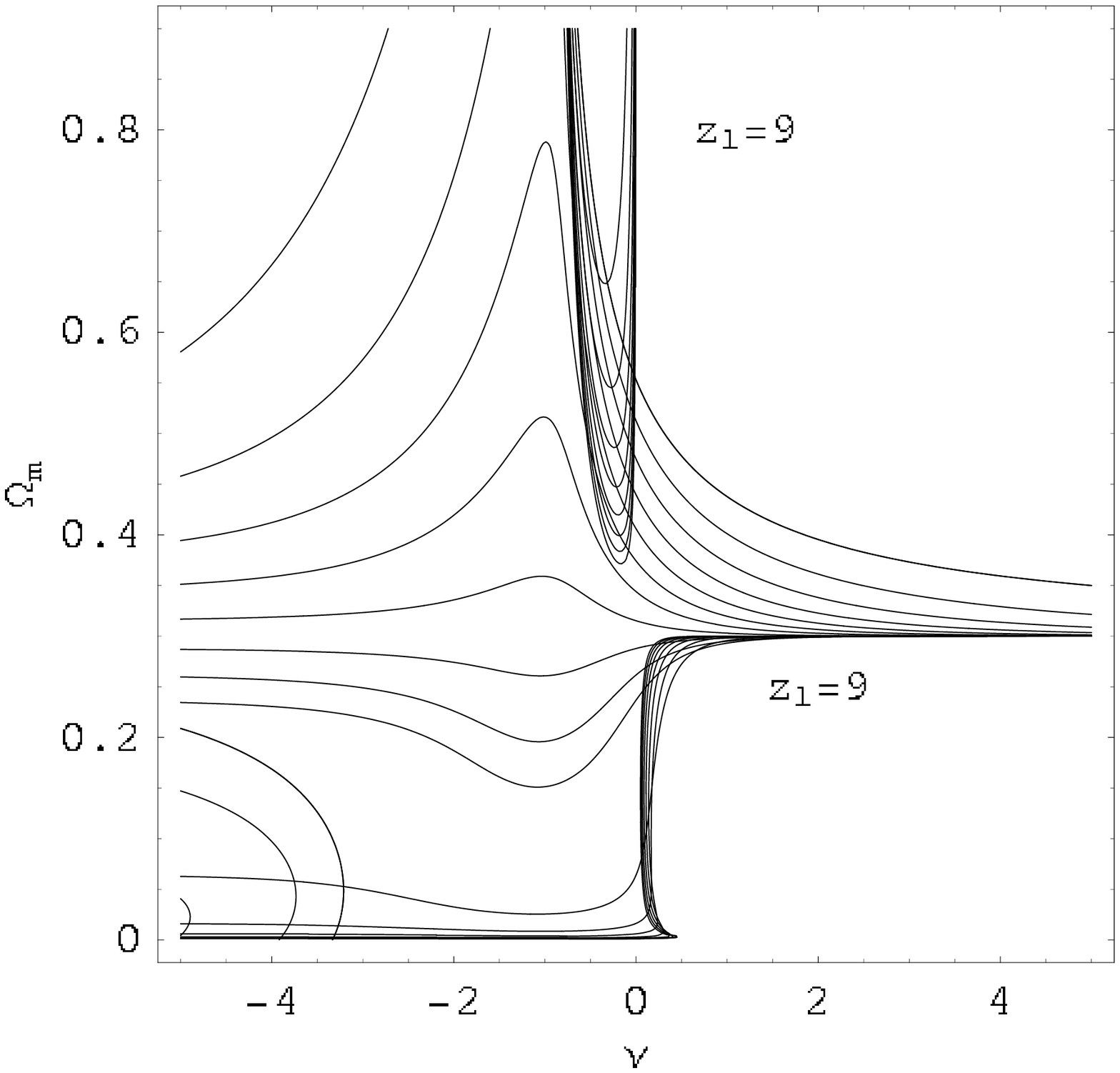}
\caption{\scriptsize{\label{cardDEzl}The loitering redshift for the Cardassian theory with CDM and dark energy in a flat universe with $\Omega_{e_0}=0.7$ and $\Gamma=0.2$. There are some similar solutions with a negative $\Gamma$ (ghost dark energy).}}
\end{figure}

\begin{figure}[h]
\centering
\includegraphics[width=6cm]{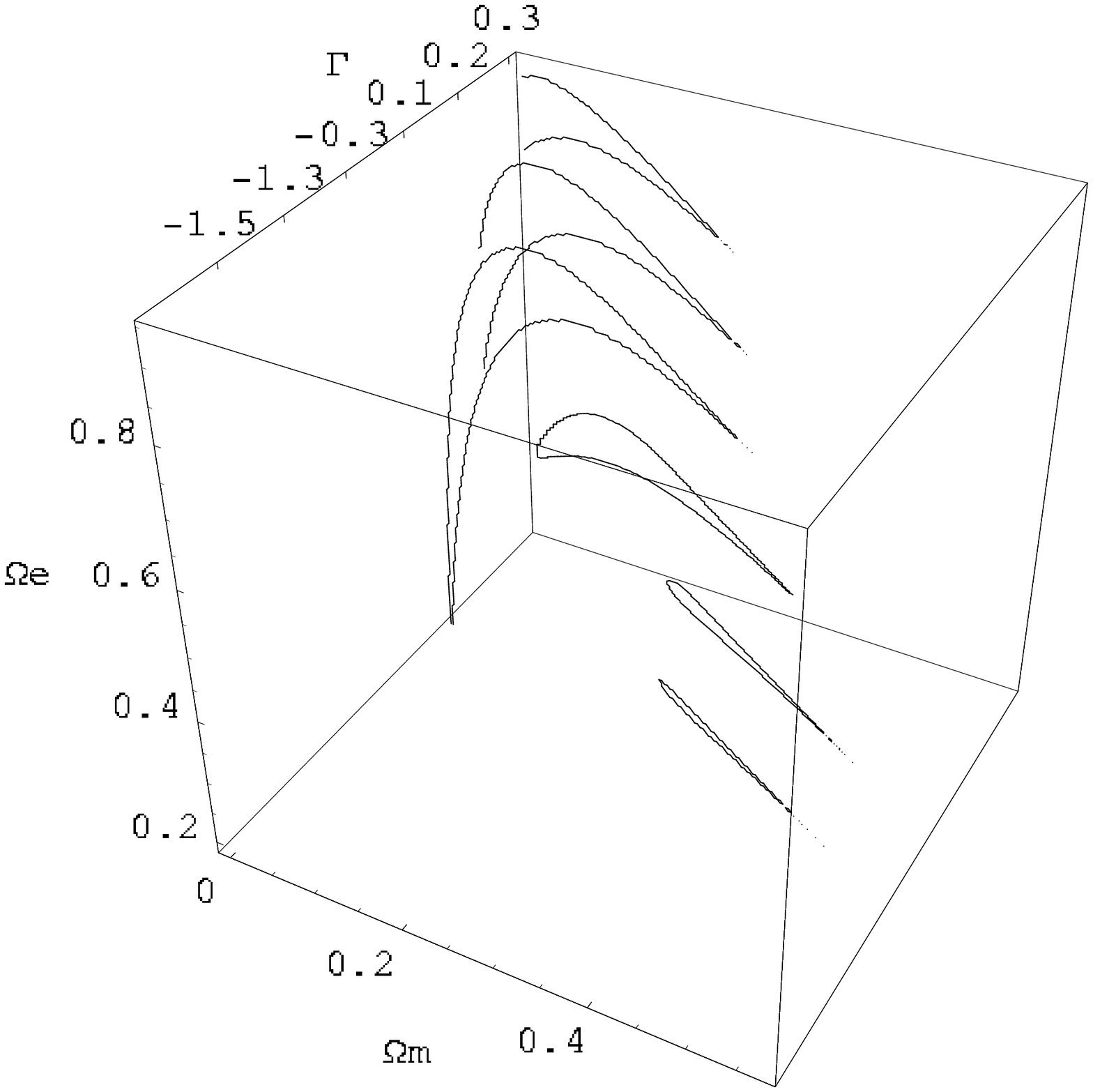}
\caption{\scriptsize{\label{cardDE}Some $2\sigma$ confidence volume of the Cardassian model with CDM and dark energy in a flat universe when $\nu=3$. It is quintessence or ghost dark energy which allows the compatibility between a positive $\nu$ and the observational data.}}
\end{figure}

Why consider dark energy on the brane when the brane alone is able to accelerate the universe expansion? If we accept the existence of branes, it seems rather logical to take into account the presence of some scalar fields predicted by high energy physics and not necessarily travelling in the bulk. When a scalar field is minimally coupled to the scalar curvature $R$ it may, in the simplest case, take the form of quintessence/ghost dark energy with a constant eos. We thus take into account the presence of such dark energy with a density $\rho_e$ in our calculation. Moreover, we show in the next section, that the Randall-Sundrum model is equivalent to the Cardassian model with dark energy $\rho_e$ having a varying eos. As usual, we begin to analyse the existence of a transient acceleration and a loitering period.\\
Some numerical solutions for the acceleration redshift $z_a$ are plotted in Fig. \ref{cardFlatDEza}. Whereas in the previous cases, $\nu$ have to be negative to allow the acceleration of the expansion, thanks to dark energy $\rho_e$ this situation changes. An accelerating universe with some reasonable values of $\Omega_{m_0}$ and $\Omega_{e_0}$ may be found with some positive values of $\nu$ and with ghost ($\Gamma<0$) or quintessence ($\Gamma>0$) dark energy.
Some curves of the Fig. \ref{cardFlatDEza} cross each other, showing the existence of some universe with transient acceleration. For instance, if we choose $(\Omega_{m_0},\Omega_{e_0},\Gamma,\nu)=(0.4,1,0,-4)$ (and thus $\Omega_{\nu_0}=-0.91$), we get an acceleration between $z=0.18$ and $z=0.53$ as shown by Fig. \ref{cardFlatDETrans}, although these values of the parameters seem unreasonable.
We will see below that some positive values of $\nu$ agree with the data when in the presence of dark energy. Thanks to this property, some early loitering epochs for reasonable values of the cosmological parameters in agreement with the data (for instance $\Omega_{m_0}\simeq 0.3$, $\Omega_{e_0}\simeq 0.7$ and $\nu\propto 0.1$) are now possible. Some solutions are displayed in Fig. \ref{cardDEzl}.
\\It is not possible to constrain this theory with the data, since there are four parameters inducing a very large degeneracy. For instance $\nu$ may be as negative as we choose. Indeed, if $\nu\rightarrow -\infty$, the brane properties of the Cardassian model disappear since we recover usual dark energy model with a constant eos, able to fit the observational data. So, one can predict that when $\nu$ tends to $-\infty$, the four dimensional $2\sigma$ confidence space $(\Omega_{m_0},\Omega_{e_0},\Gamma,\nu)$ tends to the $2\sigma$ confidence volume $(\Omega_{m_0},\Omega_{e_0},\Gamma)$ of usual dark energy model with constant eos already studied in \cite{FayFuzAli05}. In the same way, $\Gamma$ may be as negative as we want, since then the theory tends to the flat Cardassian model without any dark energy, and fits the data perfectly as shown in the previous section. Hence this model may be degenerated in at least two directions, both the negative $\nu$ and $\Gamma$.\\
An interesting property of this model is that, thanks to dark energy, some positive values of $\nu$ are now in agreement with the data. This partly explains why the Randall-Sundrum model that we will study in the next section also agrees with the data. Indeed, the Randall-Sundrum model is equivalent to the Cardassian model with $\nu=1$ and dark energy ($\rho_e$) with a varying eos (see the next section). Note that with a positive $\nu$, the WMAP prior $\Omega_{m_0}=0.27\pm 0.05$ does not hold since the Cardassian model does not tend to a linearly CDM dominated universe at large redshift as assumed by the WMAP analysis (i.e. dominated by the $\rho_m\propto (1+z)^3$ term at large $z$). Without any prior, one finds that the lowest $\chi^2$ should be around $175.26$. We do not give any values of the parameters in agreement with this last value since they are degenerated: some wide ranges of values of the parameters agree with the lowest value of $\chi^2$. A $2\sigma$ confidence volume is plotted in Fig. \ref{cardDE} with $\nu=3$. We insist that this positive value of $\nu$ is possible only thanks to the presence of dark energy that can either be quintessence or ghost. Without dark energy, only some negative $\nu$ are in agreement with the data as shown in the previous sections.
\section{Randall-Sundrum theory}\label{s3}
\subsection{Field equations}
The equations of the Randall-Sundrum type $II$ model with the dark radiation correction, a negative bulk cosmological constant and a positive brane tension write:
\begin{equation}\label{RS1}
H^2+\frac{k}{a^2}=k_1\rho_m(1+\frac{\rho_m}{2\lambda})+\frac{m}{a^4}+\frac{\Lambda}{3}
\end{equation}
\begin{equation}\label{RS2}
\dot H-\frac{k}{a^2}=-\frac{3}{2}\gamma k_1\rho_m(1+\frac{\rho_m}{\lambda})-2\frac{m}{a^4}
\end{equation}
We define the constants
\begin{itemize}
\item $\Omega_{\lambda_0}=\frac{k_1\rho_{m_0}^2}{2H_0^2\lambda}>0$, the term arising from the positive brane  tension $\lambda$ on our brane.
\item $\Omega_{d_0}=\frac{m}{H_0^2a_0^4}$, the dark radiation density parameter. $m$ is an integration constant proportional to the mass of the bulk black hole that is naked if $m<0$ (or equivalently $\Omega_{d_0}<0$).
\item $\Omega_{\Lambda_0}=\frac{\Lambda}{3H_0^2}$, the cosmological constant parameter on the brane (a contribution coming from the negative bulk cosmological constant and the brane tension).
\item $\Omega_{k_0}=\frac{-k}{H_0^2R_0^2}$, the curvature parameter.
\end{itemize}
Equations (\ref{RS1}) and (\ref{RS1}+\ref{RS2}) rewrite as
\begin{eqnarray}\label{RS3}
H^2&=&H_0^2\mbox{[}\Omega_{m_0}(1+z)^{3\gamma}+\Omega_{\lambda_0}(1+z)^{6\gamma}+\Omega_{d_0}(1+z)^4+\Omega_{\Lambda_0}+\nonumber\\
&&\Omega_{k_0}(1+z)^2\mbox{]}
\end{eqnarray}
\begin{eqnarray}\label{RS4}
\frac{1}{H_0^2}\frac{\ddot a}{a}&=&\Omega_{m_0}(1-\frac{3}{2}\gamma)(1+z)^{3\gamma}+\Omega_{\lambda_0}(1-3\gamma)(1+z)^{6\gamma}-\nonumber\\
&&\Omega_{d_0}(1+z)^4+\Omega_{\Lambda_0}
\end{eqnarray}
with $\Omega_{\lambda_0}=1-\Omega_{m_0}-\Omega_{d_0}-\Omega_{\Lambda_0}-\Omega_{k_0}$.\\
The Randall-Sundrum type $II$ model is a special case of the Cardassian theory with $\nu=1$, $A=(2\lambda)^{-1}$ and dark energy $\rho_e$ defined by a varying eos whose barotropic index would be:
$$
\gamma_e=\frac{-3k_1a^4(\lambda+\rho_m)\pm\sqrt{3}\sqrt{k_1a^4\left[6m\lambda+a^4(2\lambda\Lambda+3k_1(\lambda+\rho_m^2))\right]}}{3k_1a^4}
$$
Here $\gamma_e$ may be fully expressed as a function of the redshift by putting $\rho_m\propto a^{-3\gamma}$ and $a=a_0(1+z)^{-1}$, where $a_0$ is the scale factor today. This eos tends to the CDM eos at early-times and to the one of a cosmological constant at late times. These properties are similar to the ones of the Chaplygin gas.\\
The Randall-Sundrum type $II$ model is also equivalent to GR in four dimensions with dark energy defined by the varying barotropic index
$$
\gamma_\phi=\frac{2}{3}\frac{2\Omega_{d_0}(1+z)^4+3\gamma\Omega_{\Lambda_0}(1+z)^{6\gamma}}{\Omega_{d_0}(1+z)^4+\gamma\Omega_{\Lambda_0}(1+z)^{6\gamma}+\Omega_{\Lambda_0}}
$$
and dark energy parameter $\Omega_{\phi_0}=\Omega_{d_0}+\Omega_{\lambda_0}+\Omega_{\Lambda_0}$. At large redshift this dark energy behaves like a stiff fluid and in the future ($z\rightarrow -1$) like a cosmological constant.
\subsection{Flat universe}
\begin{figure}[h]
\centering
\includegraphics[width=4.3cm]{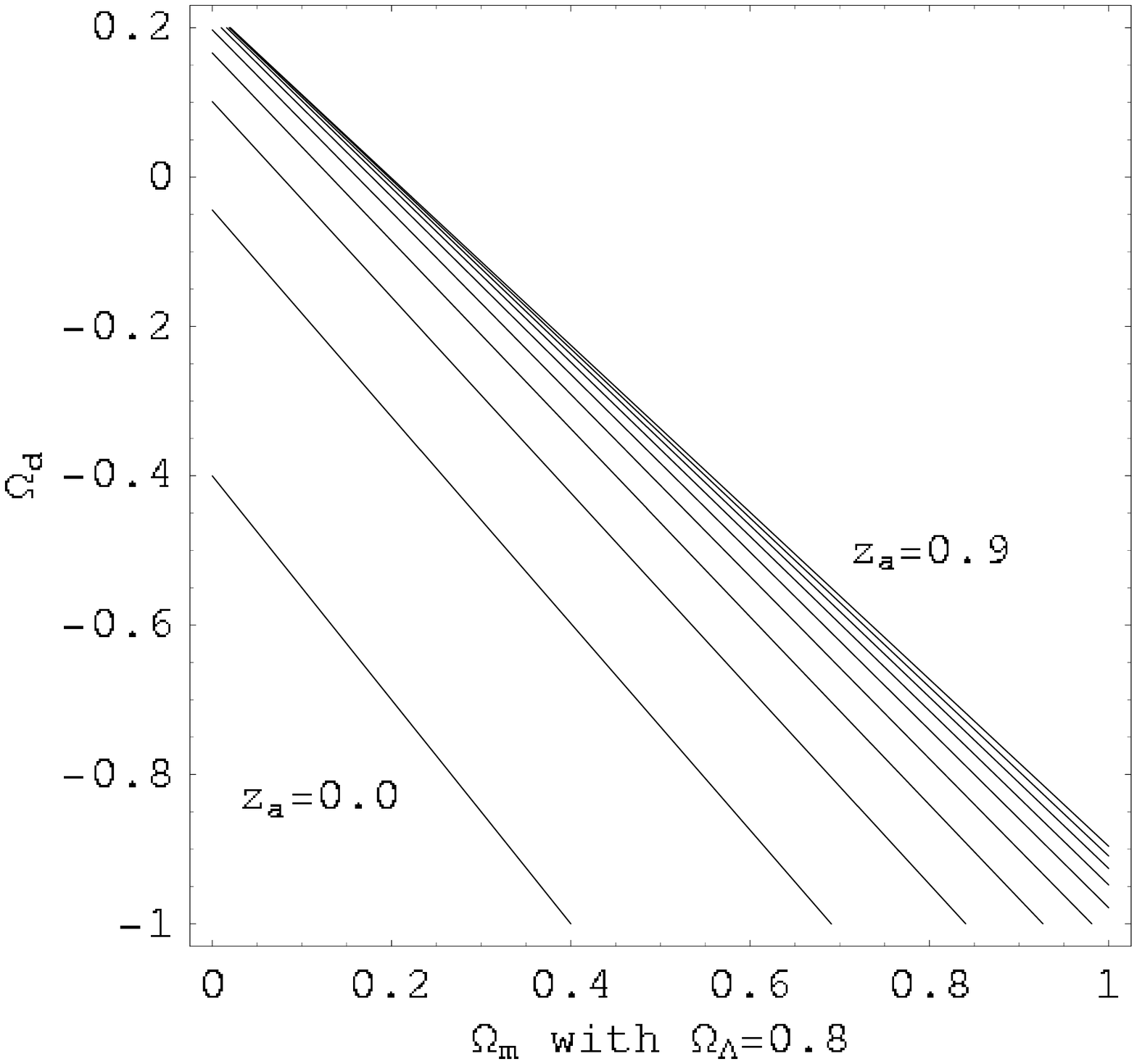}
\includegraphics[width=4.3cm]{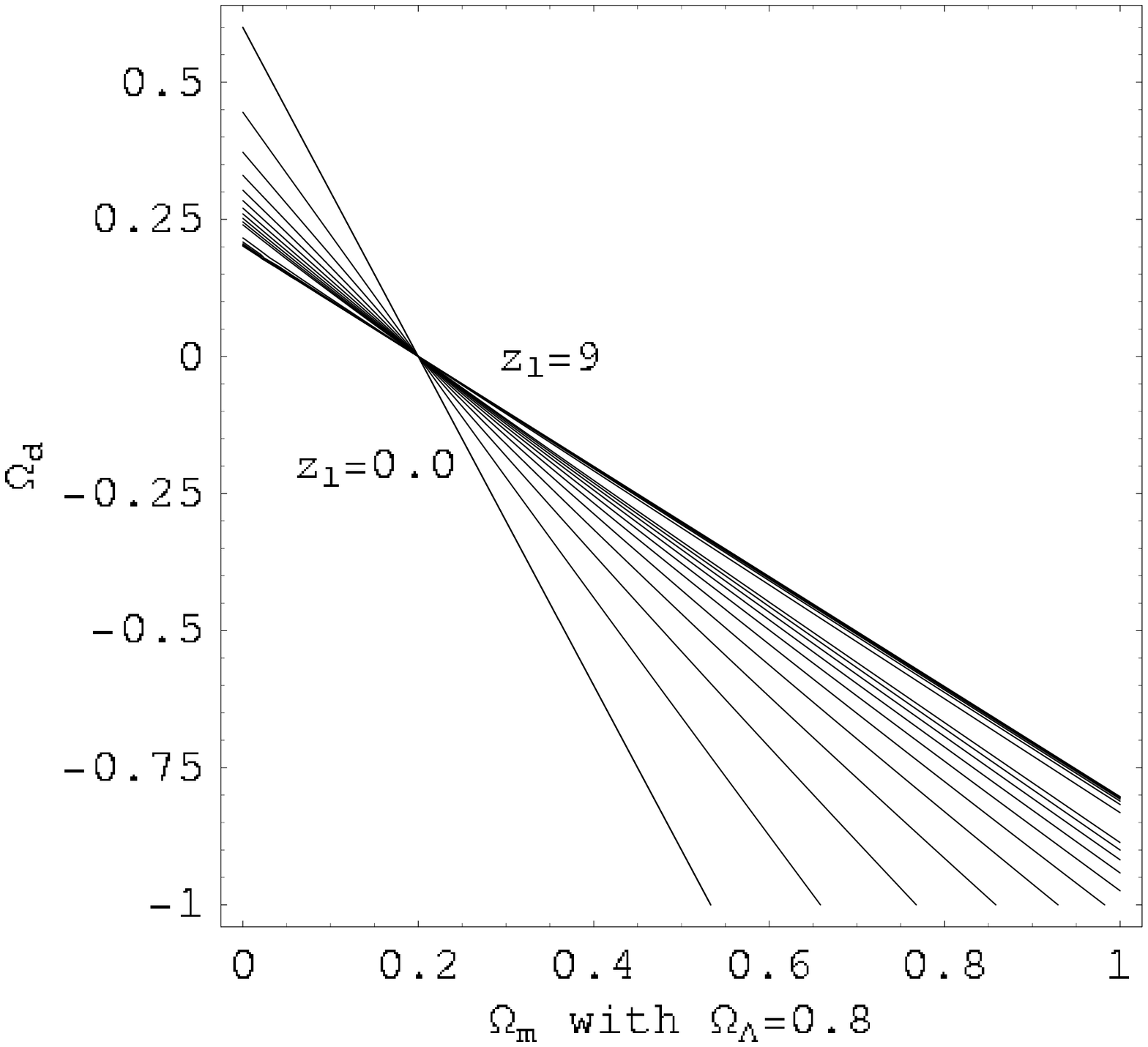}
\caption{\scriptsize{\label{RaSuFlatzazl}The acceleration (first Fig.) and loitering redshift (second Fig.) for the Randall-Sundrum theory in a flat universe when $\Omega_{\Lambda_0}=0.8$, the value giving the lowest $\chi^2$.}}
\end{figure}

\begin{figure}[h]
\centering
\includegraphics[width=6cm]{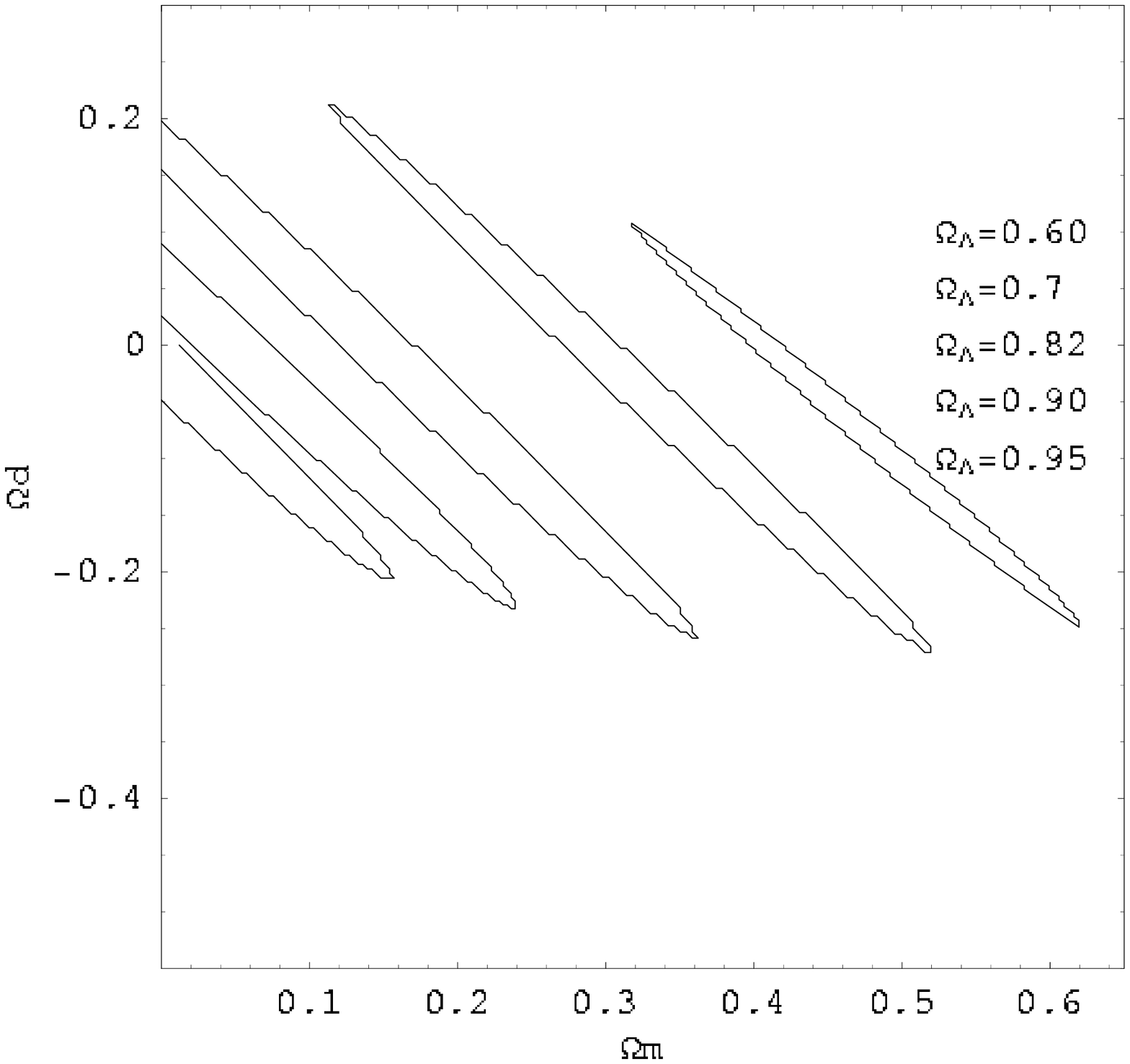}
\caption{\scriptsize{\label{RaSuFlat}$2\sigma$ contours for the Randall-Sundrum theory in a flat universe. Labels run from the right ellipse to the left one.}}
\end{figure}

Some solutions for the acceleration redshift $z_a$ are plotted on the first graph of the Fig. \ref{RaSuFlatzazl} and for the loitering redshift $z_l$ on the second graph in Fig. \ref{RaSuFlatzazl} when $\Omega_{\Lambda_0}=0.8$, the best fit value when one assumes $\Omega_{d_0}=0\pm 0.1$ (see below). A transient acceleration is possible for some smaller values of $\Omega_{\Lambda_0}$. Some loitering periods also occur.
\\
To analyse the Randall-Sundrum model with respect to the data, we need to assume some priors otherwise the lowest $\chi^2$ is got for very small $\Omega_{m_0}$ and a non negligible $\Omega_{d_0}$. Definitively, it does not look like our universe! This fact has already been noticed in \cite{Dab04} where, when neglecting $\Omega_{d_0}$, $\Omega_{m_0}=0.01$ is found. It seems dangerous to assume some usual priors coming from the CMB data because at large redshift, the $\rho_m^2$ term dominates instead of the $\rho_m$ one and the effect of dark radiation may also play an important role\citep{Gum03}. The effect of the quadratic correction $\rho_m^2/\lambda$ is negligible by the time of nucleosynthesis and thus should not act during the CMB epoch. The effect of dark radiation are analysed in \cite{Gum03}. Among others, it is shown that (quoting Gumjudpai), whereas \emph{"in the radiation era, the large scale density perturbations are suppressed at late time by a small amount, at the same time, the large scale perturbations in the dark radiation itself grow at late times"}. Today the dark radiation should be a small fraction of the radiation energy and thus, being conservative, we assume that $\Omega_{d_0}=0\pm 0.1$. Then we find $\chi^2=178.203$ with $\Omega_{m_0}=0.15$, $\Omega_{d_0}=0.008$ (the bulk black hole is not naked) and $\Omega_{\Lambda_0}=0.80$. It means that $\Omega_{\lambda_0}=0.026$, a positive value in agreement with a positive Newton constant. Note that in \cite{Dab04}, using the supernovae of \cite{Per99} and neglecting $\Omega_{d_0}$, the best $\chi^2$ is found when $\Omega_{m_0}=0.25$, $\Omega_{\Lambda_0}=0.73$ and $\Omega_{\lambda_0}=0.02$. The $2\sigma$ confidence contours are plotted in Fig. \ref{RaSuFlat} ($1\sigma$ confidence contours are too thin to be plotted). At $2\sigma$, one has $\Omega_{m_0}\in\left[0,0.62\right]$, $\Omega_{d_0}\in\left[-0.27,0.22\right]$ and $\Omega_{\Lambda_0}\in\left[0.54,1.0\right]$. The acceleration is not transient (at the limit of the $2\sigma$ contours) but an early loitering period is possible when $\Omega_{\Lambda_0}<0.8$. Minimalising $\chi^2$, we get the lowest $\chi^2$ with $H_0=64.77$. Then the universe age is $11.45$ billion years. This is too young, but higher values are possible at $2\sigma$. The acceleration of the expansion occurs in $z_a=0.44$. The linear term for the CDM dominates briefly when $0.34<z<1.17$. For larger redshift the quadratic term of the CDM dominates.
\subsection{Curved universe}
\begin{figure}[h]
\centering
\includegraphics[width=4.3cm]{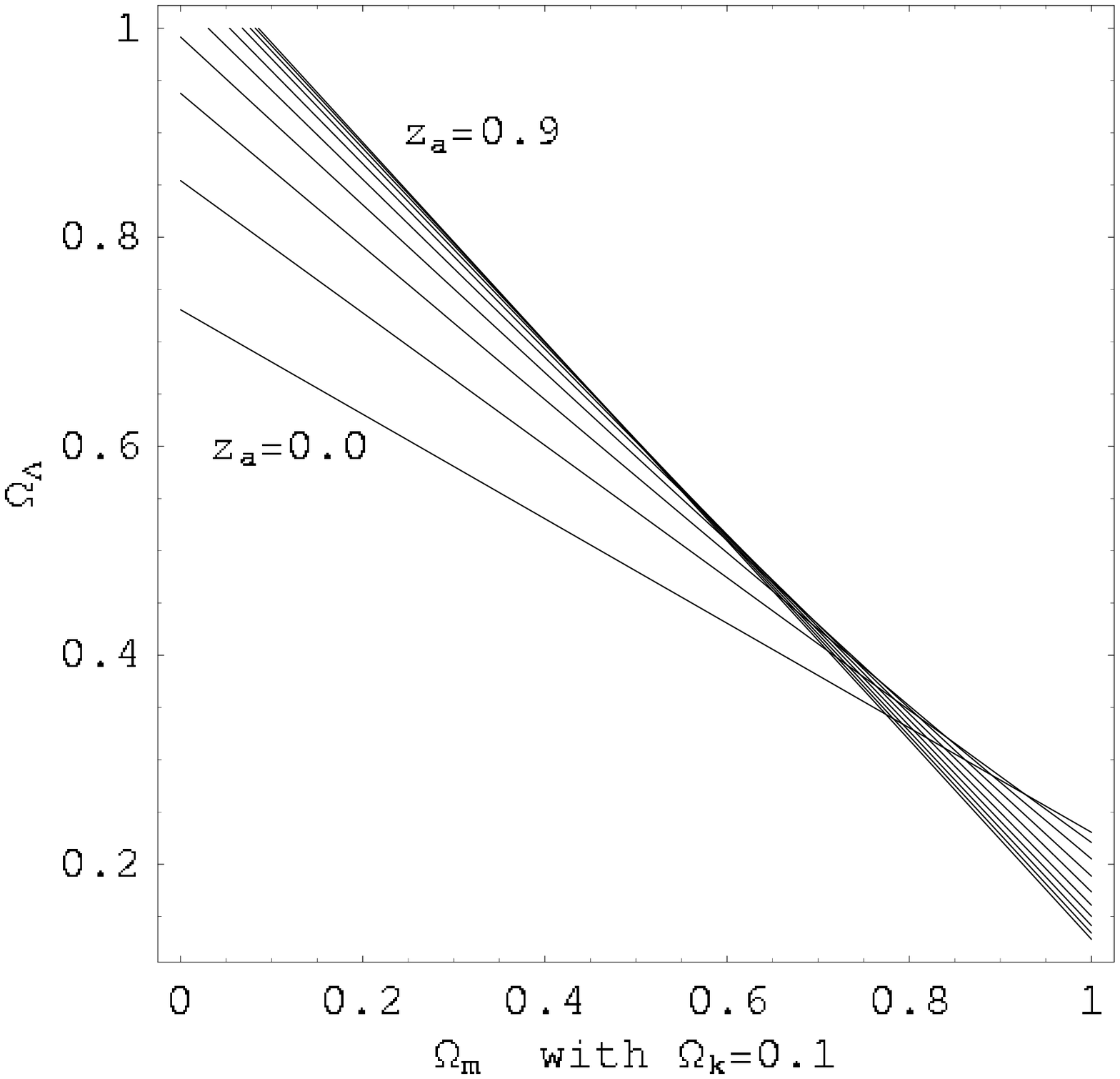}
\includegraphics[width=4.3cm]{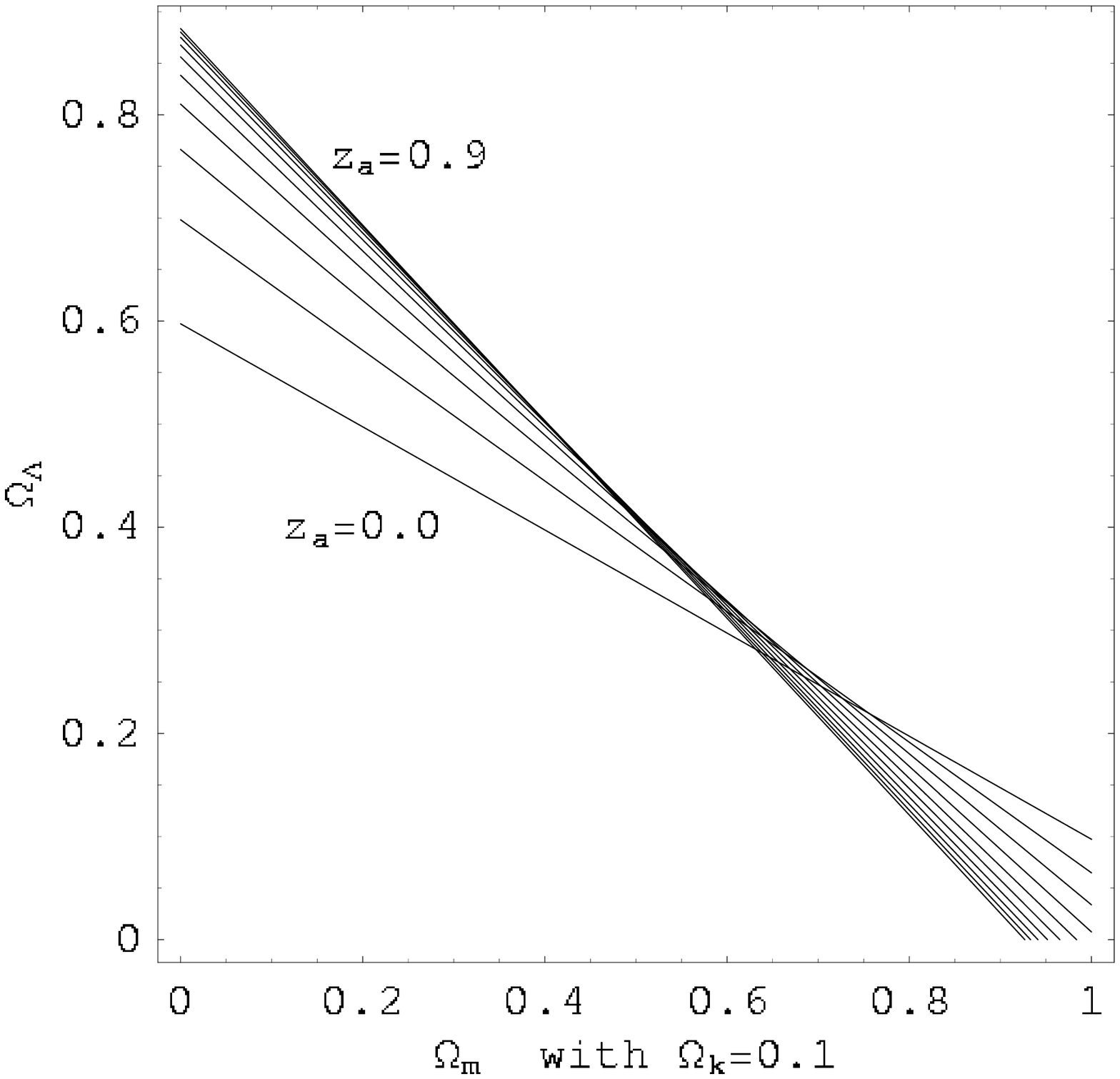}
\caption{\scriptsize{\label{RaSuCurza}Acceleration redshift $z_a$ of the Randall-Sundrum theory in a curved universe ($\Omega_{d_0}=0.08$). Some curves cross, showing the existence of a transient acceleration.}}
\end{figure}

\begin{figure}[h]
\centering
\includegraphics[width=6cm]{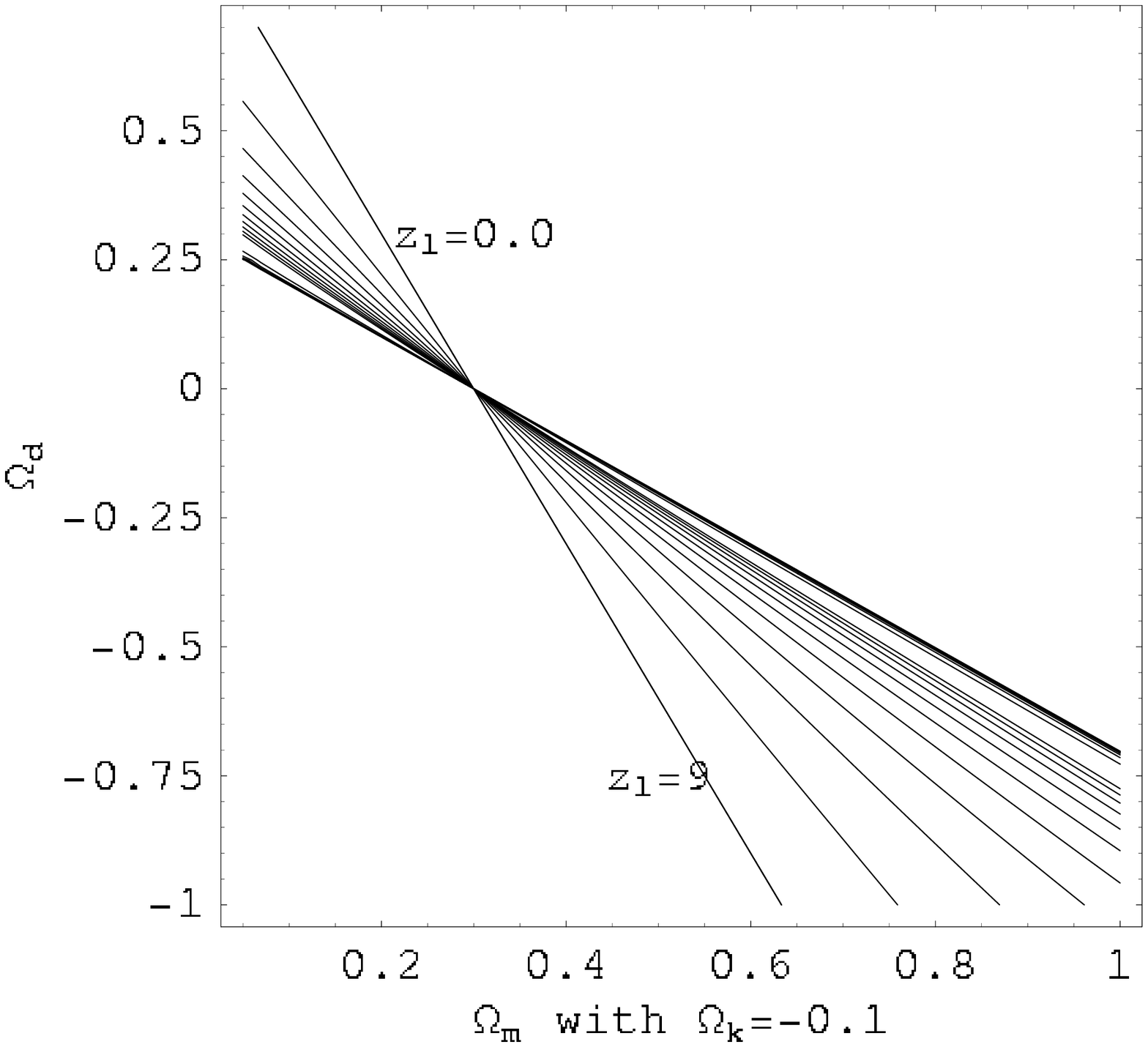}
\caption{\scriptsize{\label{RaSuCurzl}Loitering redshift $z_l$ of the Randall-Sundrum theory in a curved universe ($\Omega_{\Lambda_0}=0.8$).}}
\end{figure}

The solutions for the acceleration redshift are represented in Fig. \ref{RaSuCurza}. Some of them correspond to a transient acceleration but only for relatively low values of $\Omega_{\Lambda_0}$. These transient solutions occur for higher values of $\Omega_{m_0}$ when $\Omega_{k_0}$ becomes more and more negative. Some loitering period also exist as plotted in Fig. \ref{RaSuCurzl}.
\\When calculating the best $\chi^2$, it is really difficult to get some reasonable values for the cosmological parameters without assuming any priors for each of these parameters as shown in Table \ref{RaSuCurved}. With or without priors, the lowst $\chi^2$ is roughly the same, indicating a large degeneracy: the supernovae and CMB data are not able to constrain the Randall-Sundrum theory efficiently in a curved universe. For this reason we have not looked for the confidence contours.
\begin{table*}[!htbp]
\begin{center}
\caption{\label{RaSuCurved}\scriptsize{The lowest $\chi^2$ get for the Randall-Sundrum theory with several priors. Assuming a prior for $\Omega_{m_0}$ only gives, with respect to what we know about the universe, too a large curvature, dark radiation, and vacuum energy. The problem stays for the curvature (dark radiation) if we add a prior on dark radiation (respectively curvature).}}
\scriptsize
\begin{tabular}{llllllll}
\hline
&$\chi^2$&$\Omega_{m_0}$& $\Omega_{\Lambda_0}$ & $\Omega_{d_0}$ & $\Omega_{k_0}$&$z_a$&$H_0$\\
\hline
$\Omega_{m_0}=0.3\pm 0.1$&177.32&0.31&1.12&0.20&-0.61&$1.40<z<0.52$&64.98\\
$\Omega_{m_0}=0.3\pm 0.1$, $\Omega_{d_0}=0.0\pm 0.05$&177.98&0.30&0.89&0.004&-0.22&$z_a=0.42$&64.85\\
$\Omega_{m_0}=0.3\pm 0.1$, $\Omega_{k_0}=-0.02\pm 0.05$&178.34&0.29&0.77&-0.08&-0.02&$z_a=0.42$&64.70\\
$\Omega_{m_0}=0.3\pm 0.05$, $\Omega_{d_0}=0.0\pm 0.05$&178.58&0.29&0.76&-0.011&-0.06&$z_a=0.43$&64.54\\
and $\Omega_{k_0}=-0.05\pm 0.05$&&&&&&&\\
\hline
\end{tabular}
\end{center}
\end{table*}
\section{DGP theory}\label{s4}
\subsection{Field equations}
The field equations of the DGP theory are:
\begin{equation}\label{dgp1}
H^2+\frac{k}{a^2}=(\sqrt{k_1(\rho_m+\rho_e)+\frac{1}{4r_0^2}}\pm\frac{1}{2r_0})^2
\end{equation}
\begin{equation}\label{dgp2}
\dot H-\frac{k}{a^2}=-\frac{3}{2}k_1(\gamma\rho_m+\Gamma\rho_e)(1\pm\frac{1}{2r_0}\frac{1}{\sqrt{k_1(\rho_m+\rho_e)+\frac{1}{4r_0^2}}})
\end{equation}
where $r_0=M_4^2/(2M_5^3)$ is the cross-over scale where gravity changes its behaviour\citep{DvaGabPor00, Gum03}: on a small scale the gravitational potential behaves normally as $1/r$ but the gravity becomes $5D$ on scales larger than $r_0$. Equations (\ref{dgp1}-\ref{dgp1}+\ref{dgp2}) rewrite
\begin{eqnarray}\label{dgp3}
H^2&=&H_0^2\mbox{[}\Omega_{k_0}(1+z)^2+\mbox{[}\pm\Omega_{r_0}+\nonumber\\
&&\sqrt{\Omega_{m_0}(1+z)^{3\gamma}+\Omega_{e_0}(1+z)^{3\Gamma}+\Omega_{r_0}^2}\mbox{]}^2\mbox{]}
\end{eqnarray}
\begin{eqnarray}\label{dgp4}
\frac{1}{H_0^2}\frac{\ddot a}{a}&=&H_0^2\{\left[\pm\Omega_{r_0}+\sqrt{\Omega_{m_0}(1+z)^{3\gamma}+\Omega_{e_0}(1+z)^{3\Gamma}+\Omega_{r_0}^2}\right]^2-\nonumber\\
&&\frac{3}{2}\left[\Omega_{m_0}\gamma(1+z)^{3\gamma}+\Omega_{e_0}\Gamma(1+z)^{3\Gamma}\right]\nonumber\\
&&\left[1\pm\Omega_{r_0}\frac{1}{\sqrt{\Omega_{m_0}(1+z)^{3\gamma}+\Omega_{e_0}(1+z)^{3\Gamma}+\Omega_{r_0}^2}}\right]\}
\end{eqnarray}
with $\Omega_{r_0}=\frac{1}{2r_0H_0}$. From the constraint (\ref{dgp3}), we derive
$$
\Omega_{r_0}=\frac{\sqrt{(\Omega_{m_0}+\Omega_{e_0}+\Omega_{k_0}-1)^2}}{2\sqrt{1-\Omega_{k_0}}}
$$
The $\pm$ sign corresponds to two different ways in which the brane is embedded in the bulk\citep{Sah05}, and the cosmological consequences of these are treated in the next two sections. To simplify, we only consider the positive sign of the square root $\sqrt{x^2}=\mid x\mid$.\\
This model is equivalent to GR in four dimensions with dark energy having a varying barotropic index
$$
\gamma_\phi=\frac{\Omega_{m_0}\Omega_r\gamma(1+z)^{3\gamma}+\Omega_{e_0}\Gamma(1+z)^{3\Gamma}(\Omega_r+\sqrt{F(z)})}{\sqrt{F(z)}\left[\Omega_{e_0}(1+z)^{3\Gamma}+2\Omega_r(\Omega_r+\sqrt{F(z)})\right]}
$$
with $F(z)=\Omega_{m_0}(1+z)^{3\gamma}+\Omega_{e_0}(1+z)^{3\Gamma}\pm\Omega_r^2$. In the future when $z\rightarrow -1$, the dark energy eos tends to a $\Lambda$ eos, if $\Gamma>0$ (the brane corrections dominate), and to $\Gamma$ otherwise (ghost dark energy dominates since then $\Gamma<0$).
\subsection{The + sign}
\subsubsection{With only CDM}
\begin{figure}[h]
\centering
\includegraphics[width=4.3cm]{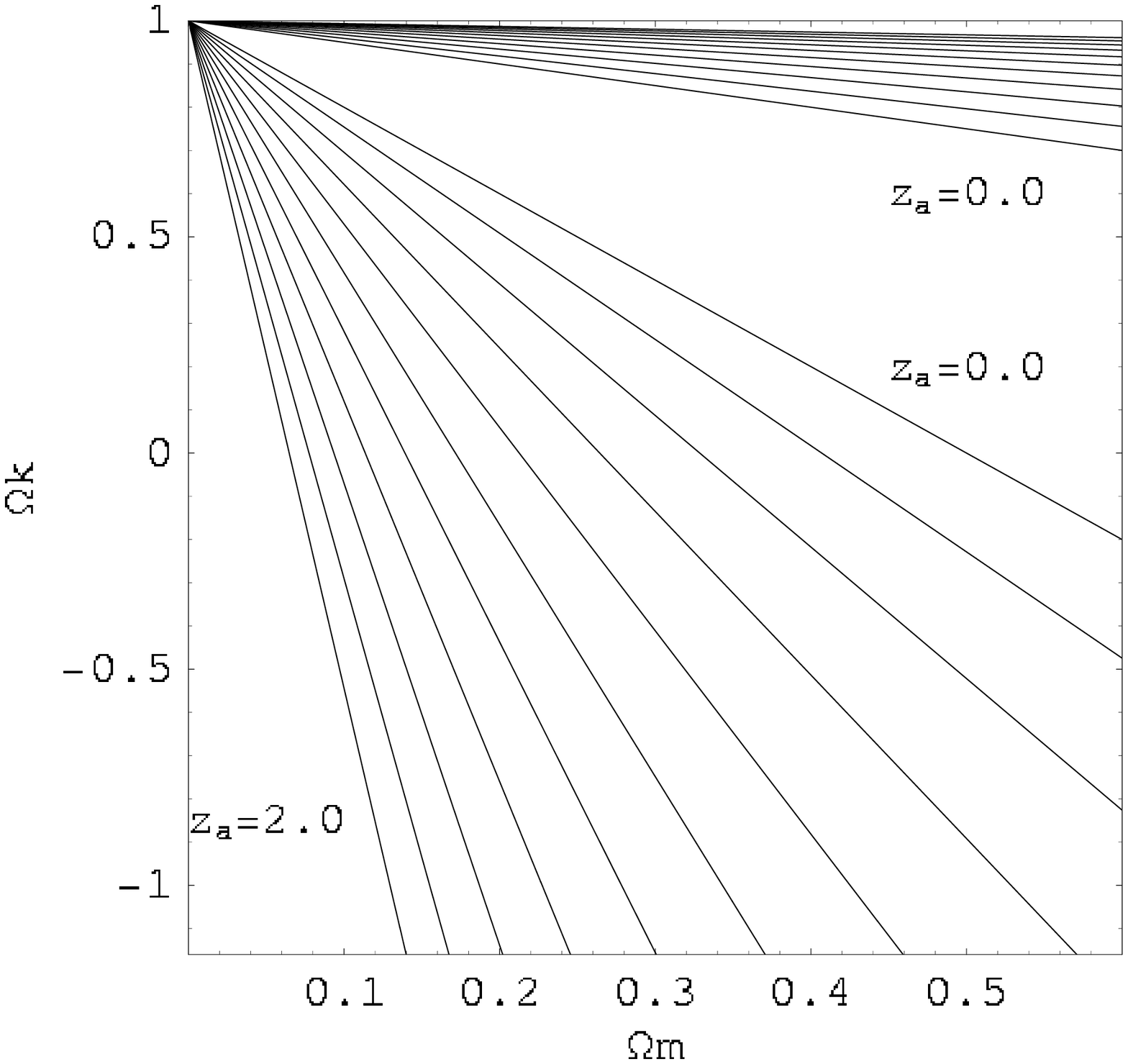}
\includegraphics[width=4.3cm]{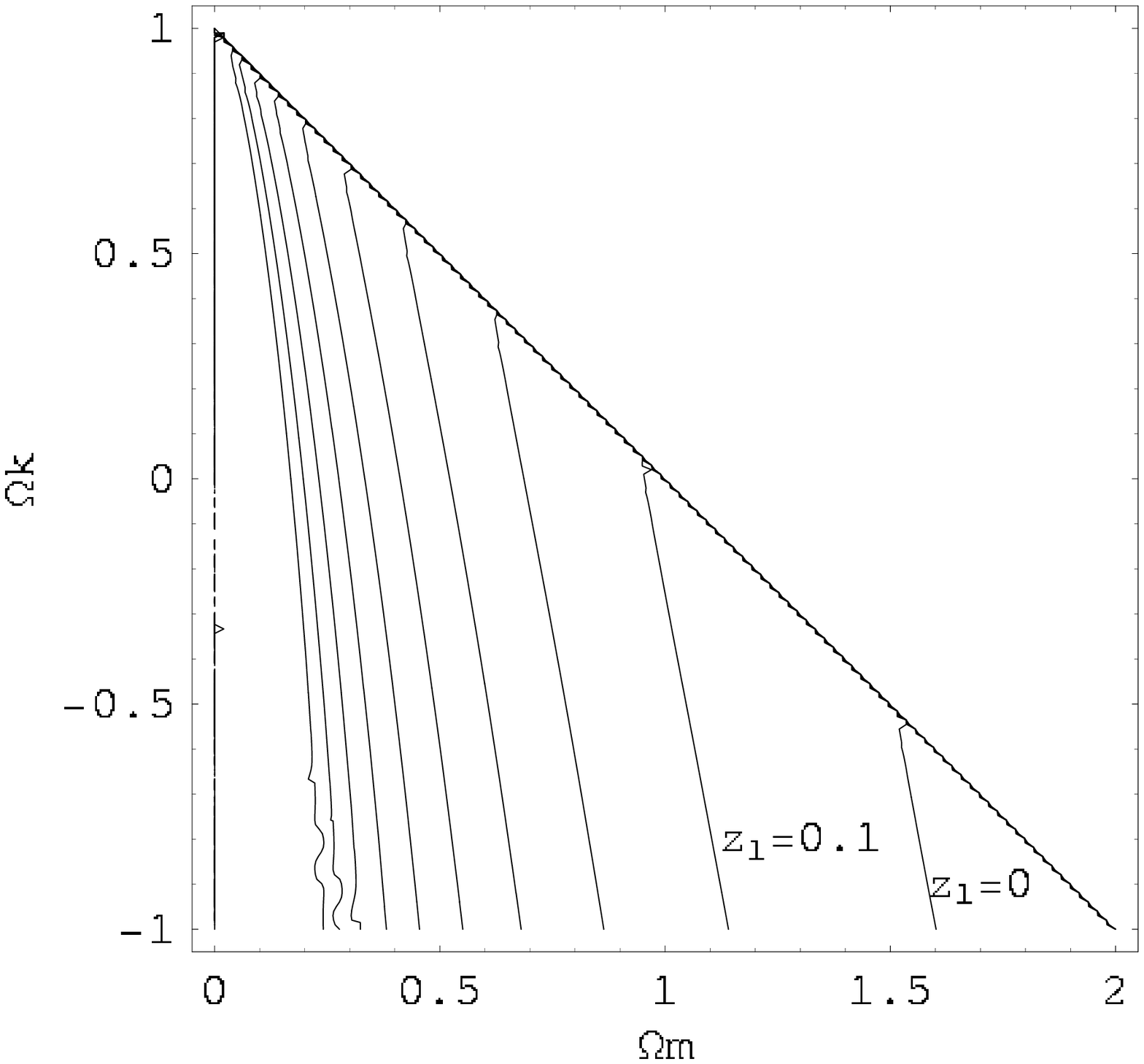}
\caption{\scriptsize{\label{DGPz}The acceleration (first Fig.) and loitering (second Fig.) redshift for the DGP theory ("+" sign) with only CDM.}}
\end{figure}

\begin{figure}[h]
\centering
\includegraphics[width=6cm]{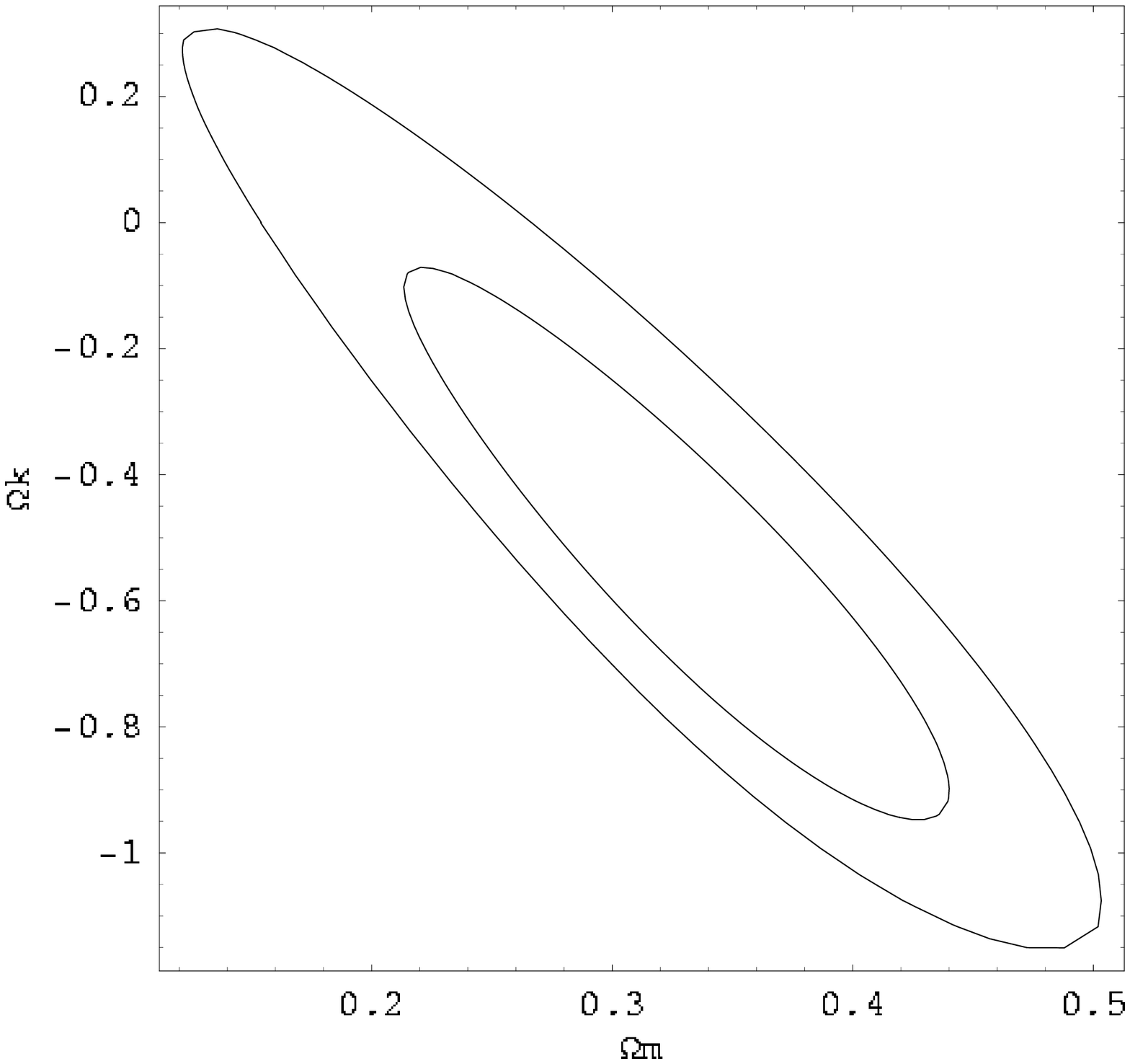}
\includegraphics[width=6cm]{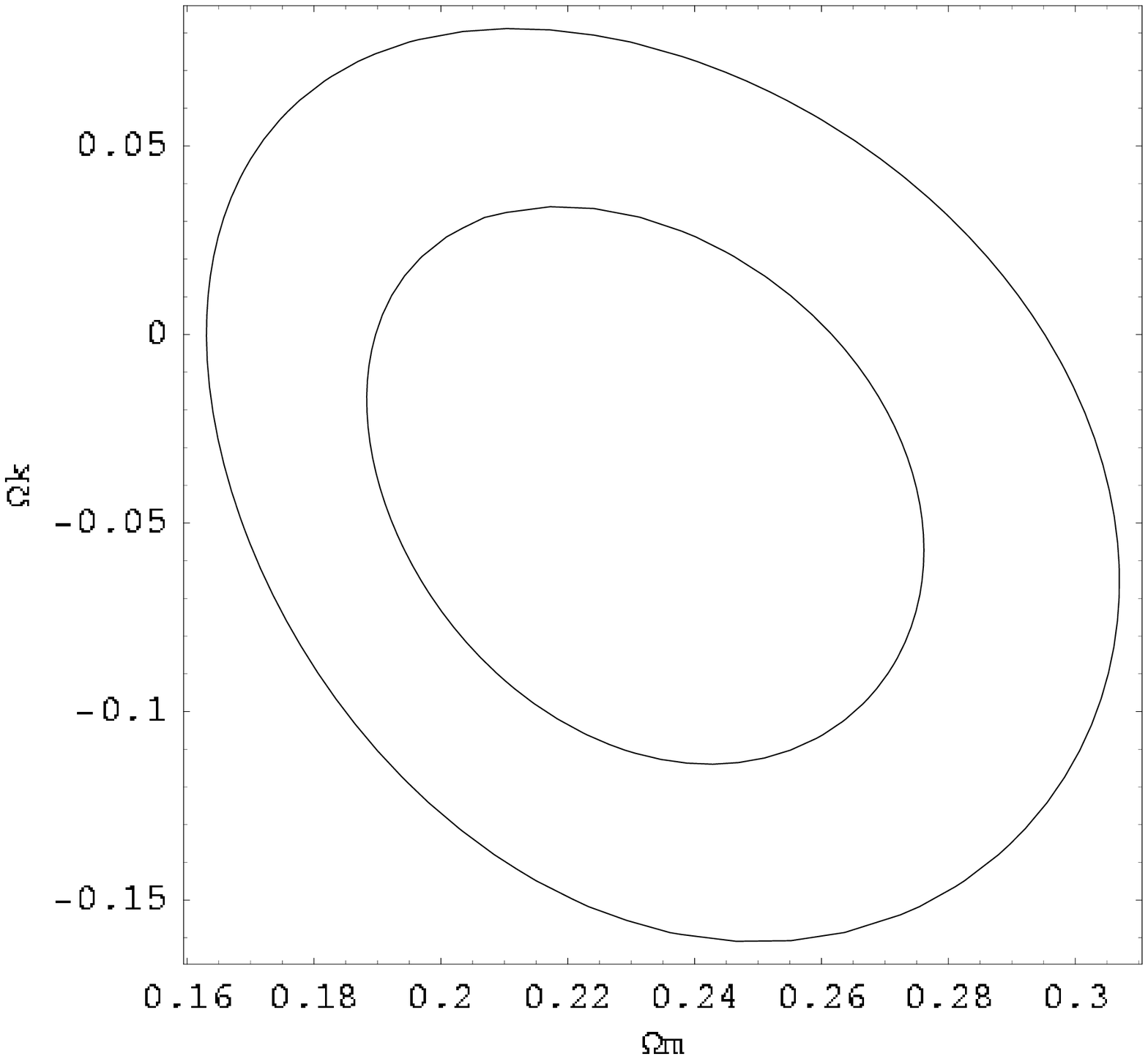}
\caption{\scriptsize{\label{DGPCurve}The confidence contours of the DGP theory ("+" sign) with CDM. The first graph used the supernovae data alone whereas the second one also takes into account the WMAP priors $\Omega_{m_0}=0.27\pm 0.05$ and $\Omega_{k_0}=-0.02\pm 0.05$.}}
\end{figure}

The acceleration redshift $z_a$ without any dark energy is written as
\begin{equation}
z_a=\left[4\frac{\Omega_r^2(3\gamma-1)}{\Omega_{m_0}(2-3\gamma)^2}\right]^{\frac{1}{3\gamma}}-1.
\end{equation}
Some solutions are plotted on the first graph in Fig. \ref{DGPz}. The curves do not cross each other, so the expansion acceleration is not a transient phenomenon. All the curves correspond to an expansion that is both presently accelerating and following a decelerated expansion. There is no analytical expression for the loitering redshift $z_l$, but some numerical solutions are shown on the second graph of Fig. \ref{DGPz}.
\\
When we only use the supernovae data, the lowest $\chi^2$ is $\chi^2=177.846$ with $\Omega_{m_0}=0.33$ and $\Omega_{k_0}=-0.56$ in agreement with \cite{AlcPir04}. The $1\sigma$ and $2\sigma$ confidence contours are plotted on the first graph in Fig. \ref{DGPCurve}. At $2\sigma$, one has $\Omega_{m_0}\in\left[0.12,0.50\right]$ and $\Omega_{k_0}\in\left[-1.16,0.32\right]$. After minimalisation, the Hubble constant is $64.86$ and the universe is $15.35$ billion years old. The expansion accelerates for $z_a=0.80$.\\
The value of $\Omega_{k_0}$ for the lowest $\chi^2$ seems really too high. Hence, we assume the WMAP priors $\Omega_{m_0}=0.27\pm 0.05$ and $\Omega_{k_0}=-0.02\pm 0.05$. Since at large $z$, the linear CDM term dominates, it should be reasonable. Then, the best $\chi^2$ is $\chi^2=181.46$ ($\chi^2_{DOF}=1.17$) with $\Omega_{m_0}=0.23$ and $\Omega_{k_0}=-0.04$. If the value of the curvature parameter is reasonable, $\Omega_{m_0}$ seems low with respect to $WMAP$ data. The confidence contours are plotted in the second graph in Fig. \ref{DGPCurve}. At $2\sigma$, one has $\Omega_{m_0}\in\left[0.16,0.30\right]$ and $\Omega_{k_0}\in\left[-0.16,0.08\right]$. After minimalisation, the Hubble constant is $H_0=64.02$ and the age of the universe is $14.96$ billions years. The acceleration redshift is $z_a=0.76$ and the domination of the CDM occurs for $z_d=0.68$, justifying that at early times the CDM dominates.\\
The DGP model with CDM seems to favour a closed universe. A loitering period is possible at late times, i.e. $z_l<1.5$. Concerning the cross over scale $r_0$, we found with the supernovae data and the CMB priors that $\Omega_{r_0}\in\left[0.35,0.44\right]$, that is $r_0H_0\in\left[1.13,1.42\right]$. With the value of the Hubble constant in megaparsecs, it means that $r_0\simeq 6$ gigaparsecs. This is in good agreement with \cite{DefLan02} where, with $52$ supernovae\cite{Rie99} and some CMB data in a flat universe, $r_0H_0\simeq 1.4$.
\subsubsection{Flat universe with dark energy}
\begin{figure}[h]
\centering
\includegraphics[width=4.3cm]{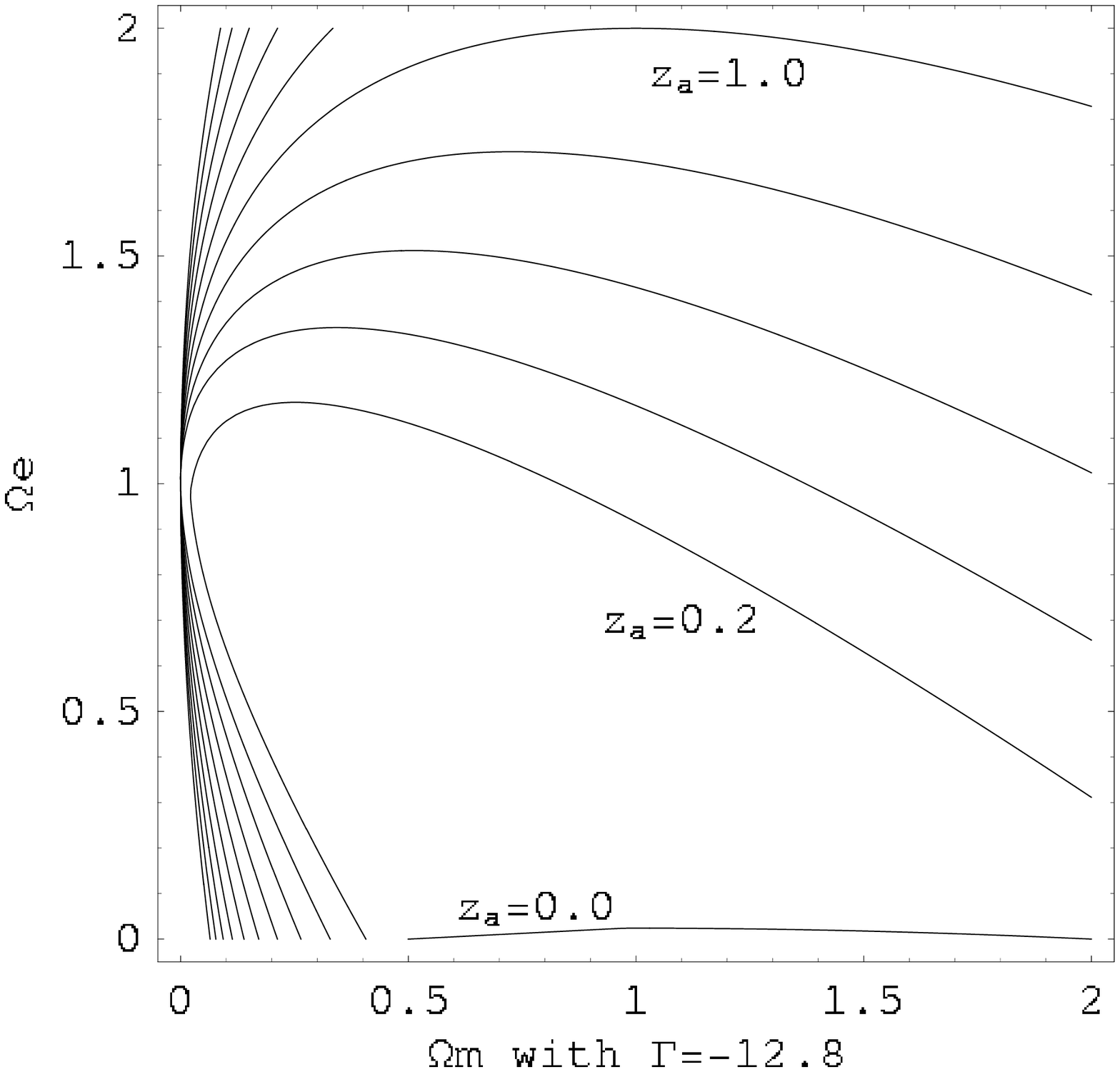}
\includegraphics[width=4.3cm]{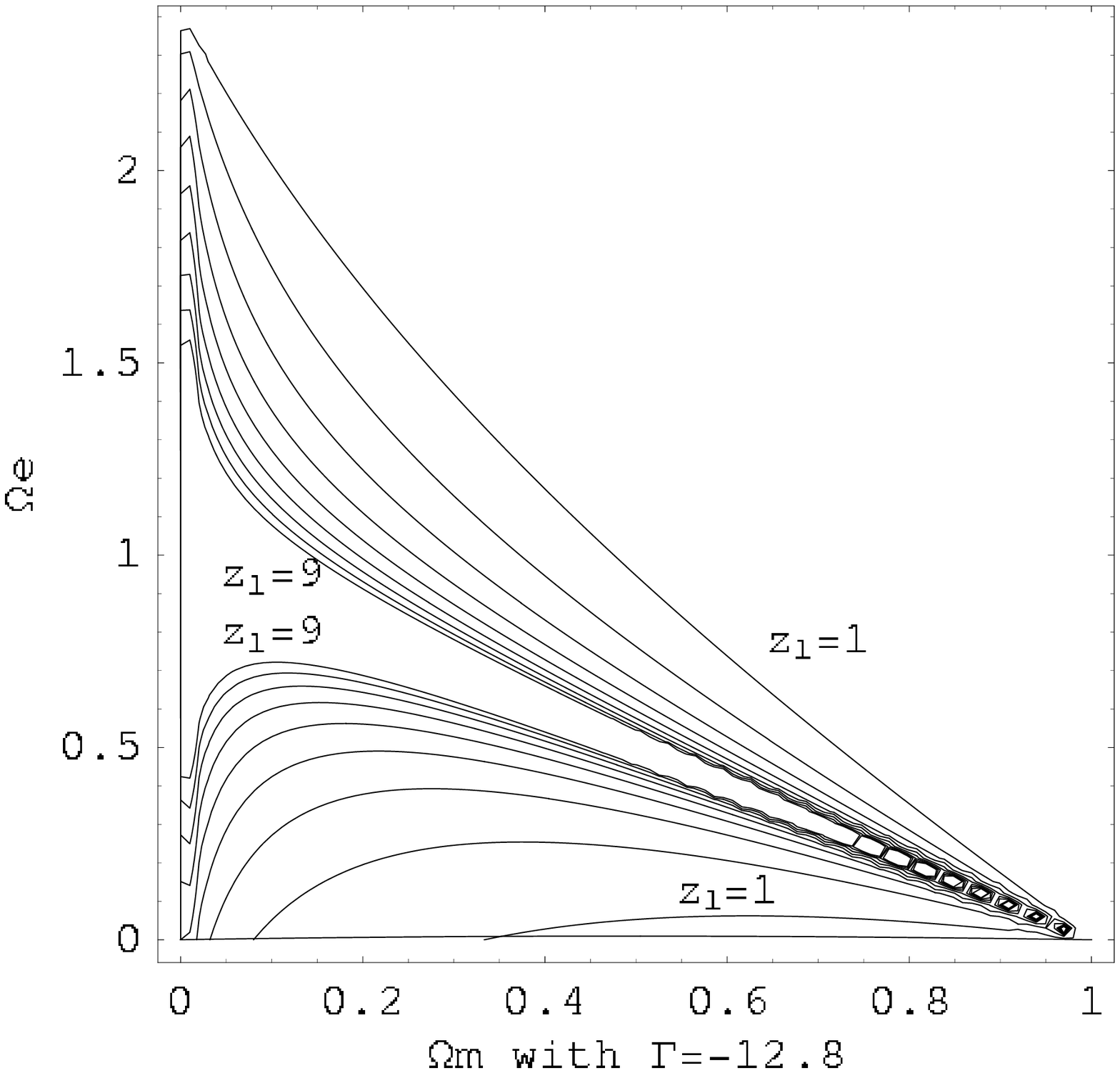}
\caption{\scriptsize{\label{DGPDEza}The acceleration and loitering redshift for the DGP theory ("+" sign) with dark energy on the brane.}}
\end{figure}

\begin{figure}[h]
\centering
\includegraphics[width=9cm]{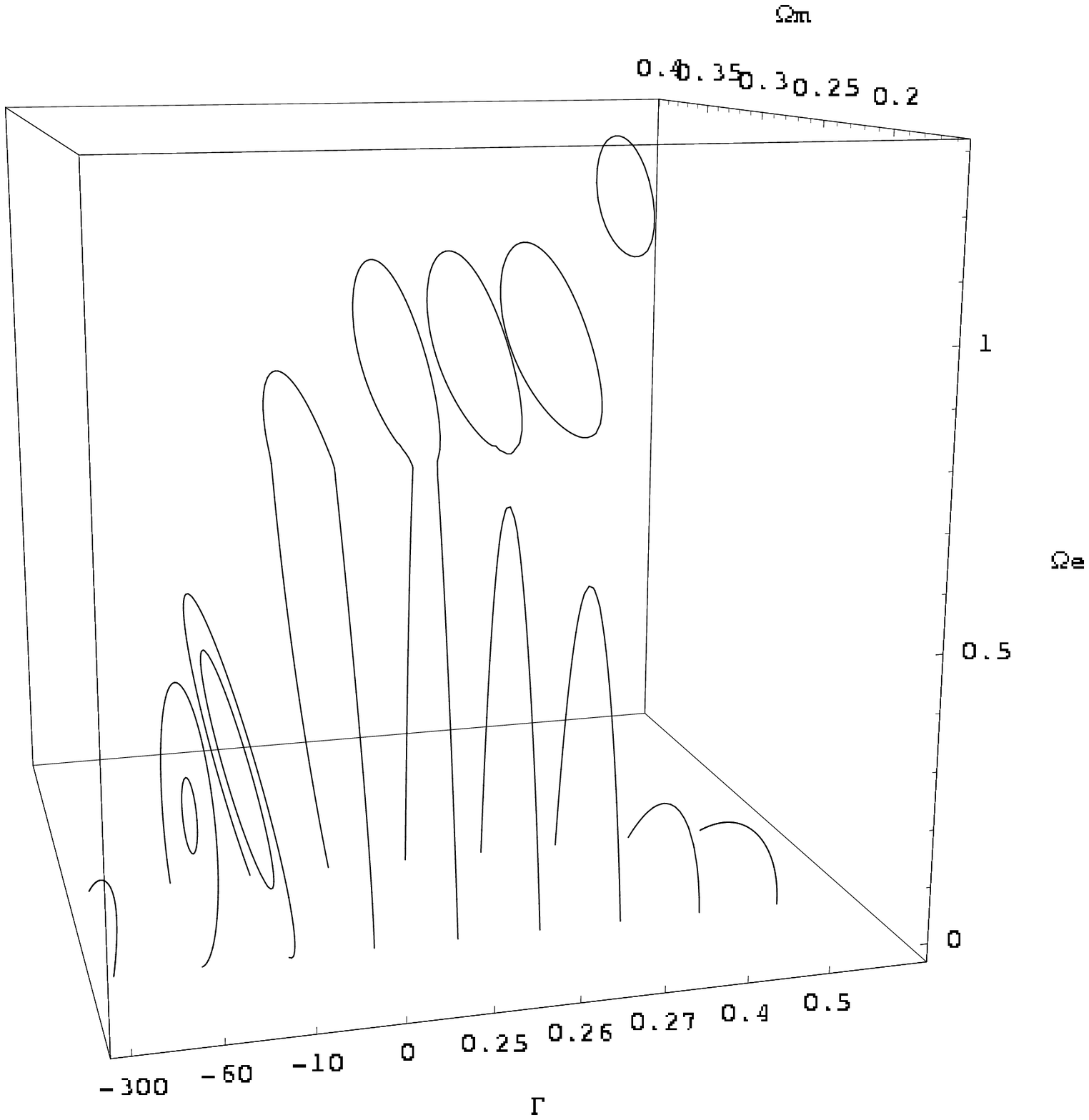}
\caption{\scriptsize{\label{DGPDE}The confidence contours of the DGP theory ("+" sign) with dark energy in a flat universe.}}
\end{figure}

Some solutions for the acceleration and loitering redshift are plotted on the Fig. \ref{DGPDEza}. We have not detected any transient acceleration for reasonable values of $\Omega_{m_0}$ or $\Omega_{e_0}$ but loitering occurs.
\\Using the supernovae data alone, we find $\chi^2=175.32$ for the lowest $\chi^2$ with $\Omega_{m_0}=0.30$, $\Omega_{e_0}=0.18$ and $\Gamma=-9.76$. This time, $\Omega_{m_0}$ is close to the WMAP value $0.27$. The $2\sigma$ confidence contours are very large in particular for the $\Gamma$ parameter which may be as negative as we choose. This is rather logical. The DGP model does not need any dark energy to reproduce the observed acceleration and then, dark energy may be as low as necessary, that is, with $\Gamma\rightarrow -\infty$ or $\Gamma\rightarrow +\infty$ with $\Omega_{e_0}\rightarrow 0$. The other parameters are also very degenerated. At $2\sigma$ one has $\Omega_{m_0}\in\left[0,25.9\right]$ and $\Omega_{e_0}\in\left[0,83\right]$.\\
To reduce these confidence contours, we assume the $WMAP$ prior $\Omega_{m_0}=0.27\pm 0.05$. It should be justified as long as $\Gamma<\gamma=1$ (for the CDM) because then the CDM dominates at early-times. Then the best fit is found for $\chi^2=175.39$ with $\Omega_{m_0}=0.27$, $\Omega_{e_0}=0.17$, and $\Gamma=-12.83$. These results are qualitatively the same as without the prior. However, the confidence contours are smaller but for the parameter $\Gamma$, which is still strongly degenerated for the same reason as before\footnote{caution: when one considers some low values of $\Gamma$ with some high values of $\Omega_{e_0}$, the universe is not necessarily CDM dominated at large redshift and the WMAP prior may be invalidated}. The confidence contours are represented in the Fig. \ref{DGPDE}. At $2\sigma$, one has $\Omega_{m_0}\in\left[0.16,0.4\right]$ and $\Omega_{e_0}\in\left[0,1.45\right]$. Minimalisation gives $H_0=67.75$ and the universe age is then $14.13$ billion years. The acceleration of the expansion occurs for $z_a=0.32$ and the universe starts being CDM dominated when $z_d=0.33$. Concerning the cross over scale $r_0$, we found with the supernovae data and the CMB priors that $\Omega_{r_0}\in\left[0.008,0.4\right]$, that is, $r_0H_0\in\left[1.25,62\right]$. With the value of the Hubble constant in megaparsec, it means that $5<r_0 <\simeq280$ gigaparsecs. The cross-over scale is thus very degenerated and can be larger than without dark energy.
early loitering agrees with the data if $\Gamma$ is small ($O(10^{-1})$).
\subsection{The - sign}
\subsubsection{With only CDM}
\begin{figure}[h]
\centering
\includegraphics[width=12cm]{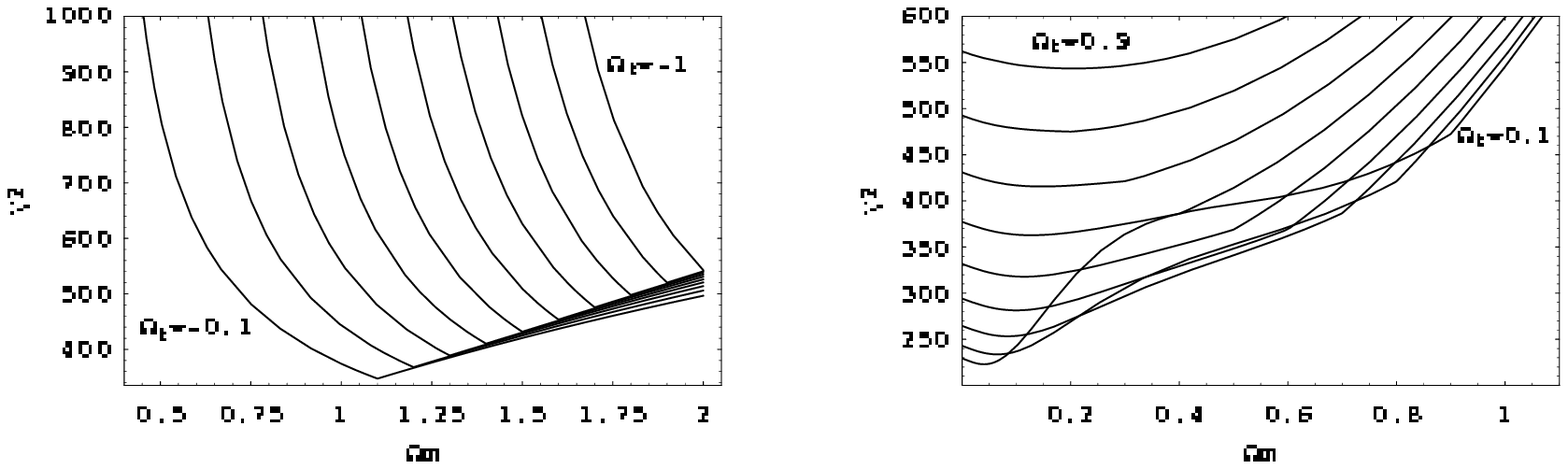}
\caption{\scriptsize{\label{DGPNeg}The lowest $\chi^2$ for the DGP theory ("-" sign) in a universe with CDM. It is plotted as a function of $\Omega_{m_0}$ for different values of $\Omega_{k_0}$. Clearly, this model does not fit the data.}}
\end{figure}

This model does not fit the supernovae data. This is shown in Fig. \ref{DGPNeg} where we have plotted the lowest $\chi^2$ for some values of $\Omega_{m_0}$ and $\Omega_{k_0}$.
Thus this brane model is excluded by the observations. Things are different if we add dark energy as shown in the next section.
\subsubsection{Flat universe with dark energy}
\begin{figure}[h]
\centering
\includegraphics[width=12cm]{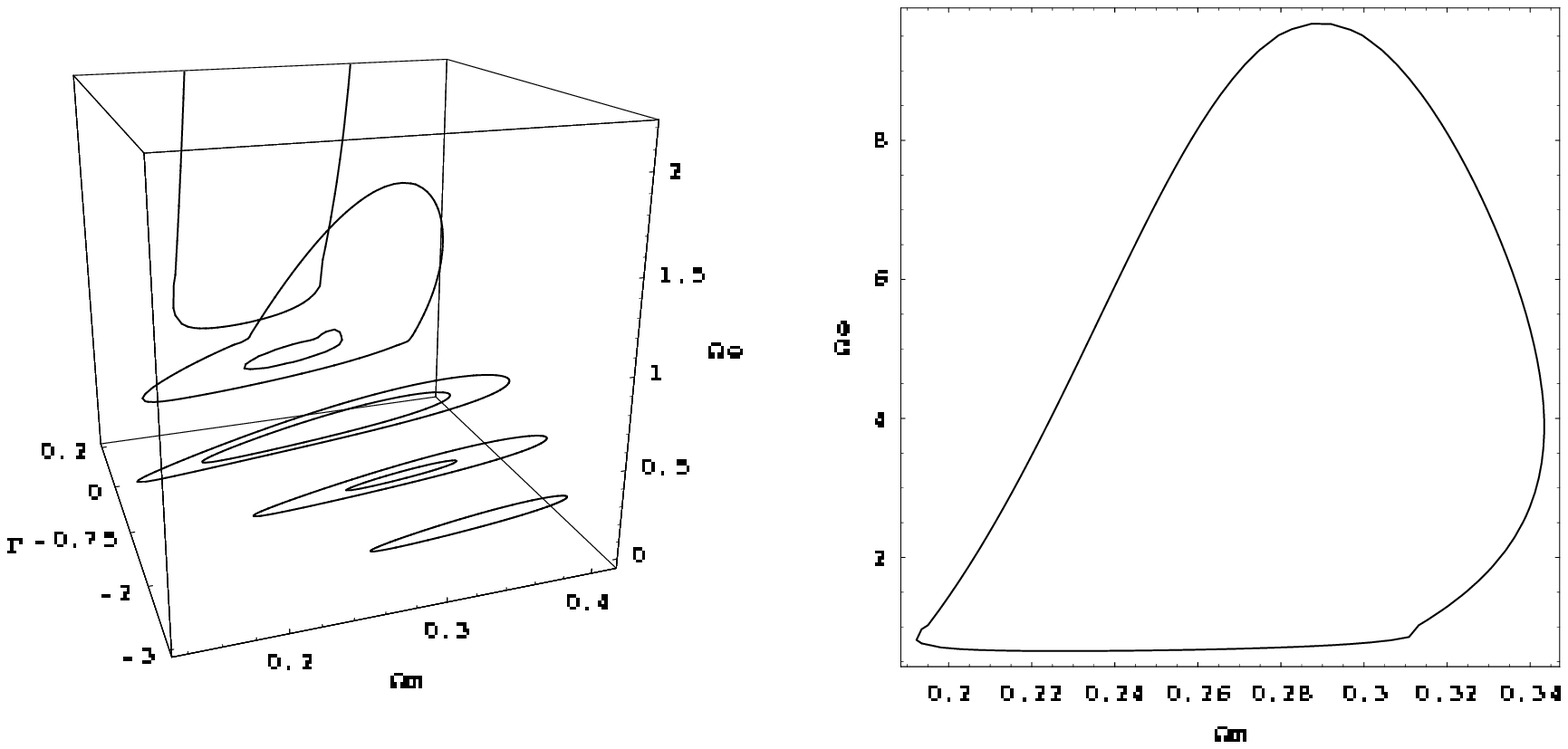}
\caption{\scriptsize{\label{DGPNegCMB}$1\sigma$ and $2\sigma$ confidence volume for the DGP theory ("-" sign) with dark energy in a flat universe when we assume the prior $\Omega_{m_0}=0.27\pm 0.05$. The second graph shows the $2\sigma$ interval when $\Gamma=0.3$.}}
\end{figure}

Adding dark energy on the brane allows us to reconcile this model with the observations. The acceleration redshift is plotted in the first Fig. (\ref{DGPDEMinusza}) when $\Gamma=-0.75$, the value matching the lowest $\chi^2$ (see below). Some similar Fig.s may be got with some other values of $\Gamma$. There is no transient acceleration but a loitering epoch may occur as shown by the right graph in Fig. \ref{DGPDEMinusza}.
\begin{figure}[h]
\centering
\includegraphics[width=4.3cm]{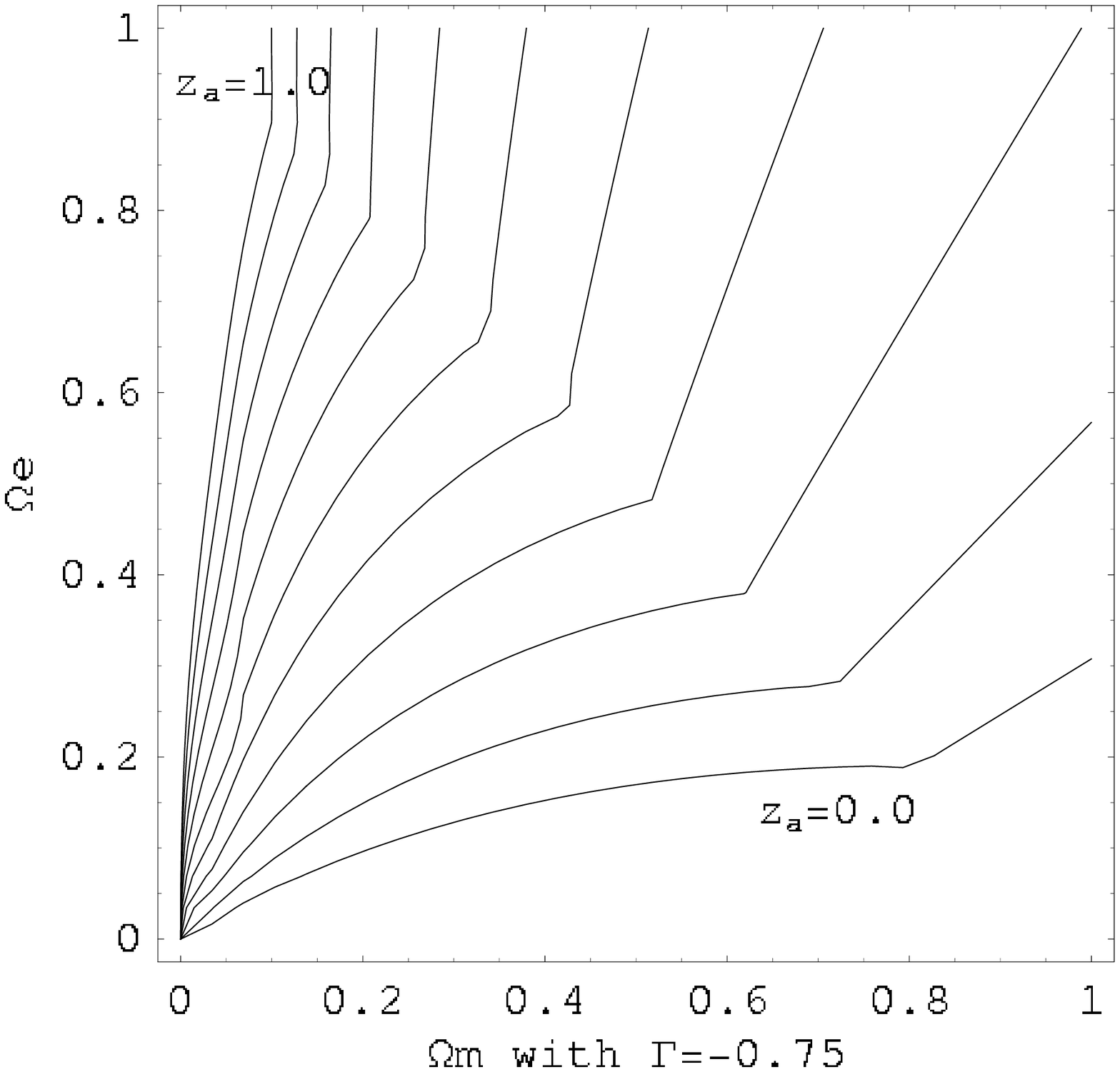}
\includegraphics[width=4.3cm]{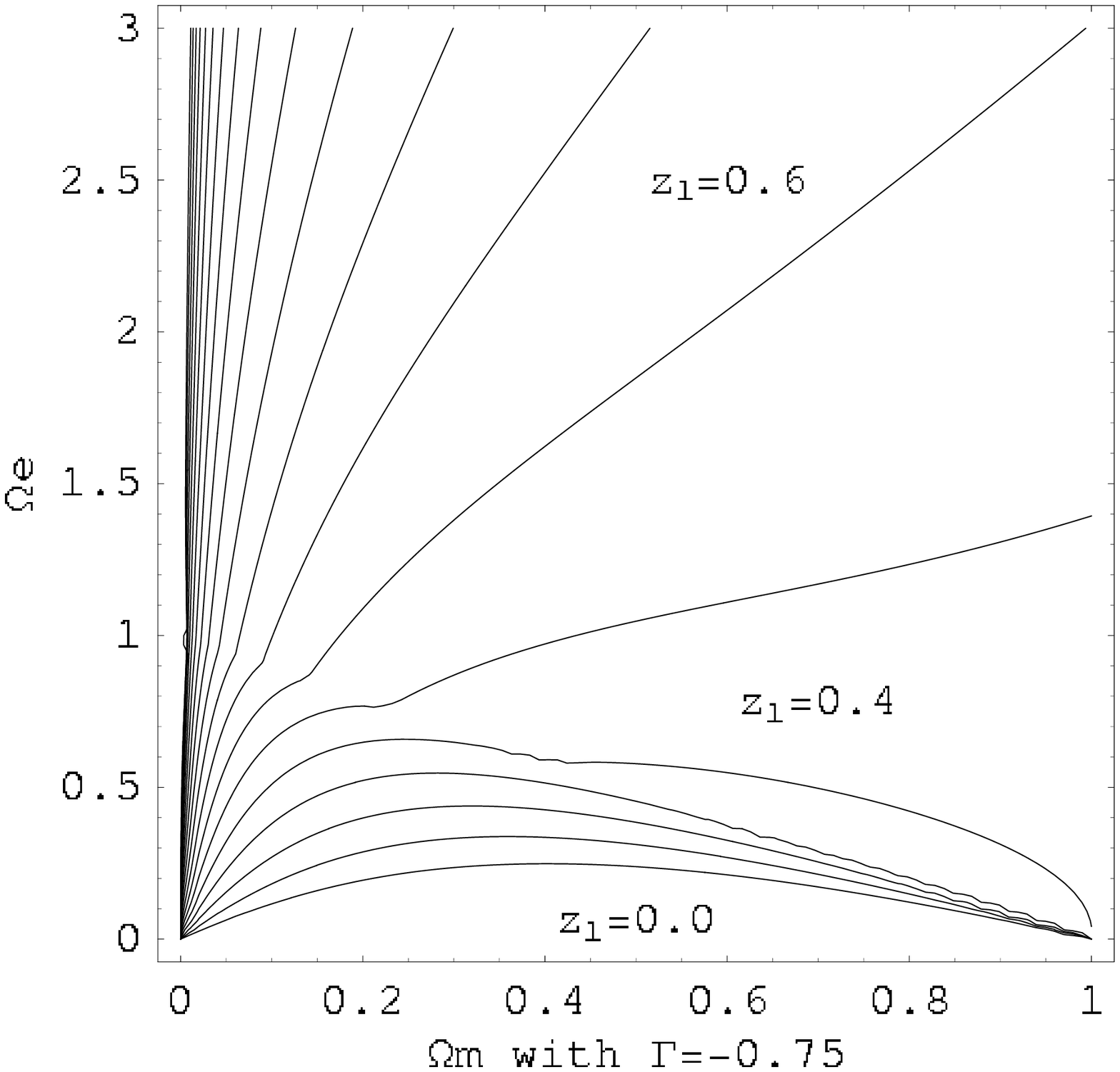}
\caption{\scriptsize{\label{DGPDEMinusza}Acceleration and loitering redshift for the DGP theory ("-" sign) with dark energy and in a flat universe when $\Gamma=-0.75$, the values of $\Gamma$ matching the lowest $\chi^2$.}}
\end{figure}
\\
Without any prior, we get $\chi^2=176.40$ with $\Omega_{m_0}=0.49$, $\Omega_{e_0}=0.51$ and $\Gamma=-1.38$. The cosmological parameters are highly degenerated. At $2\sigma$, one has $\Gamma\in\left[-11,0.41\right]$. This is a smaller interval than with the "+" sign. It comes from the fact that now dark energy is necessary to fit the data and thus $\Gamma$ cannot take any value. However $\Omega_{e_0}$ may be larger than $280$ and $\Omega_{m_0}$ larger than $120$! After minimalisation, the lowest $\chi^2$ is got with $H_0=66.15$. The age of the universe is then $13.17$ billion years. The universe begins to accelerate when $z_a=0.29$ and is dominated very recently by the CDM, since $z_d=0.005$.\\
Now we assume the WMAP prior $\Omega_{m_0}=0.27\pm 0.05$. Then, we get $\chi^2=176.74$ with $\Omega_{m_0}=0.27$, $\Omega_{e_0}=0.40$, and $\Gamma=-0.75$. The $2\sigma$ confidence contours are plotted in Fig. \ref{DGPNegCMB}. One has $\Gamma\in\left[-3.95,0.29\right]$, $\Omega_{m_0}\in\left[0.14,0.41\right]$, $\Omega_{e_0}\in\left[0.23,20.8\right]$. After Minimalisation, one finds $H_0=96.47$ and the age of the universe is $13.06$ billion years. The acceleration starts when $z_a=0.33$ and the linear term of the CDM, $\rho_m\propto (1+z)^3$, will soon stop dominating the universe since $z_d=-0.05$. This value of $H_0$ seems physically unreasonable. However, at $1\sigma$, it is possible to build a DGP model with a reasonable value of the Hubble constant. For instance, if we impose $H_0=65$ and we minimalise the $\chi^2$, we find $\chi^2=177.38$ with $\Omega_{m_0}=0.30$, $\Omega_{e_0}=0.69$ and $\Gamma=-0.01$. Thus we conclude that the DGP model with the minus sign and dark energy is in agreement with the data and the usual value of the Hubble constant only if this model is very close to a $\Lambda CDM$ model. Mathematically, it induces a very small (large) $\Omega_{r_0}$($r_0$). The vanishing of $\Omega_{r_0}$ means then that the cross over scale $r_0$ for which the gravity becomes $5D$ tends to infinity when the DGP model becomes indistinguishable from a $\Lambda CDM$ one.
An early loitering period agrees with the data for the positive values of $\Gamma$.
\section{Conclusion}\label{s5}
\begin{figure*}[h]
\centering
\includegraphics[width=12cm]{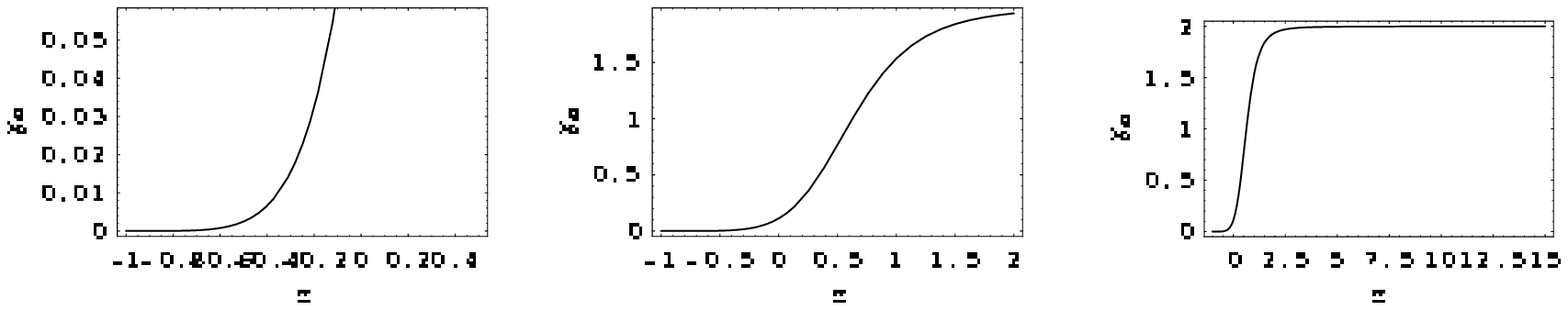}
\caption{\scriptsize{\label{DarkRS1}The Randall - Sundrum theory is equivalent to GR with dark energy. These curves represent the barotropic index of dark energy for a Randall - Sundrum model with only CDM and for several intervals of the redshift when one assumes $\Omega_{d_0}= 0\pm 0.1$.}}
\end{figure*}

\begin{figure*}
\centering
\includegraphics[width=12cm]{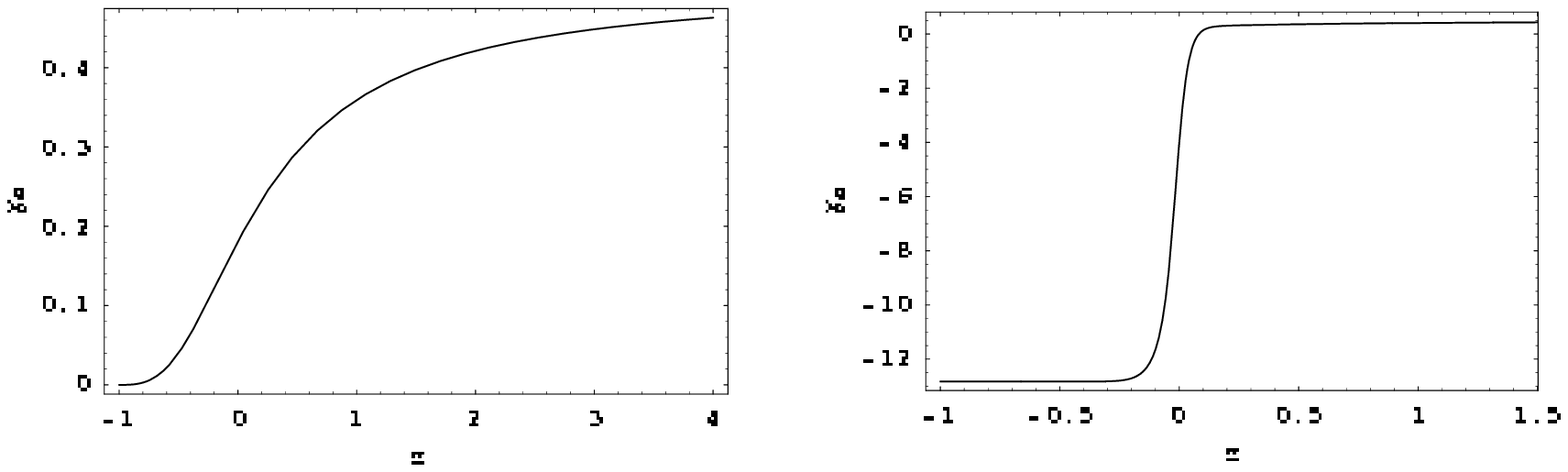}
\caption{\scriptsize{\label{DarkDGP}The DGP theory is equivalent to GR with dark energy. These curves represent the barotropic index of this dark energy when one considers the RS with only CDM (left graph) or with dark energy on the brane (right graph).}}
\end{figure*}

The parameter values corresponding to the lowest $\chi^2$ for each brane theory in agreement with and reasonably constrained by both the supernovae and CMB data are given in Tab. \ref{tabChi2A}-\ref{tabChi2B}. These are the models we are going to discuss now.\\
\\
The Cardassian theory with only CDM is equivalent to GR with dark energy $\rho_\phi$ with a constant eos very close to the one of a cosmological constant $p_\phi/\rho_\phi=-1$: for a flat universe, the barotropic index $\gamma_\phi=p_\phi/\rho_\phi+1=-0.01$ whereas with curvature, $\gamma_\phi=0.02$. Mathematically, there is no transient acceleration. A late loitering period agrees with the data. Adding dark energy does not allow us to fit the data better, but in this case the $\nu$ parameter may take some positive values that otherwise disagree with an accelerated expansion.\\
The Randall-Sundrum theory illustrates how the conclusions got with the data strongly depends on the assumption we made on the cosmological parameters.\\
First, we assume that $\Omega_{d_0}= 0\pm 0.1$. The Randall-Sundrum theory is equivalent to the Cardassian theory or to GR, both with dark energy having a varying eos. In this last case, the barotropic index of dark energy mimicking the Randall - Sundrum theory with only CDM is represented in Fig. \ref{DarkRS1}. In the past, the barotropic index tends to a stiff fluid, and in the future to a cosmological constant. A transcient acceleration disagrees with the data (at the border of the $2\sigma$ confidence contours) but an early loitering period may occur.\\
If we decide to change the prior and take, for instance, $\Omega_{m_0}=0.30\pm 0.1$, something new happens. The best $\chi^2$ is got when $\Omega_{m_0}=0.29$, $\Omega_{\Lambda_0}=0.78$, and $\Omega_{d_0}=-0.09$. A loitering period is always in agreement with the data, and a transient acceleration always at the border of the $2\sigma$ confidence contours. But this dark energy eos mimicking the Randall-Sundrum model and shown on the Fig. \label{DarkRS} presents some very interesting properties. In the past, the barotropic index tends to a stiff fluid. Later, it mimics quintessence dark energy and then ghost dark energy: dark energy is thus able to cross the line $\gamma_\phi=0$. In the far future, it will tend to a cosmological constant. This crossing of the line $\gamma_\phi=0$ has also been noticed in \cite{AreKos05}. Sometimes, one wonders what the physical meaning of ghost dark energy is that violates the weak energy condition or what it means when quintessence dark energy becomes ghost. Here, we see that these questions could be physically meaningless if we try to answer in the framework of dark energy. However, they become physically meaningful if we consider that quintessence/ghost dark energy is just a mathematical representation of a physically well justified theory such as the Randall-Sundrum theory.\\
\begin{figure*}[h]
\centering
\includegraphics[width=12cm]{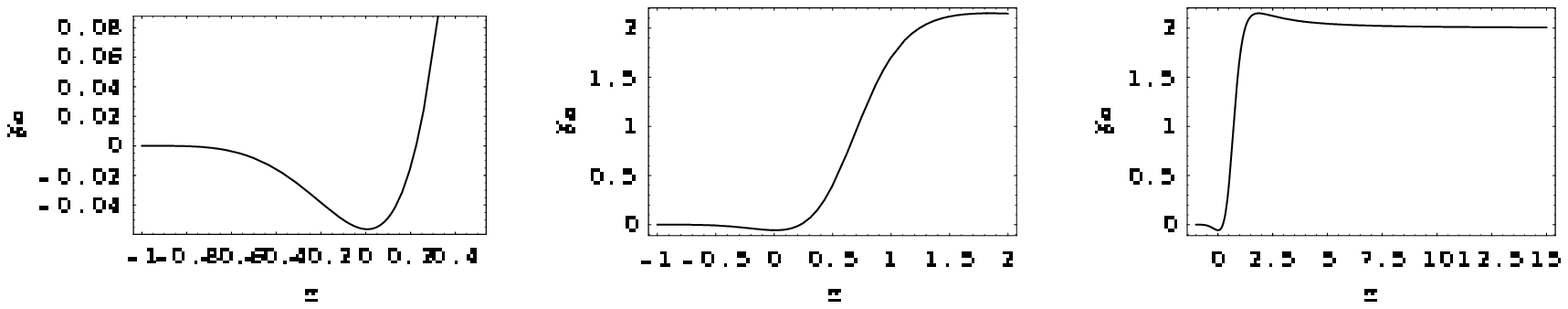}
\caption{\scriptsize{\label{DarkRS}The Randall - Sundrum theory is equivalent to GR with dark energy. These curves represent the barotropic index of this dark energy for a Randall - Sundrum model with only CDM and for several intervals of the redshift when one assumes $\Omega_{m_0}=0.30\pm 0.1$.}}
\end{figure*}
What about the DGP theory? It is also equivalent to GR with dark energy. The barotropic index corresponding to the DGP theory with only CDM or with dark energy on the brane is represented in Fig. \ref{DarkDGP}. In the first case, dark energy is quintessence. It tends to a cosmological constant in the future and to the special value $\gamma_\phi=0.5$ in the past. In the second case, dark energy is a ghost and tends to a cosmological constant in the past and to $\Gamma$ in the future.\\
For the DGP theory with only CDM, a transient acceleration is mathematically impossible and a late loitering period ($z_l<1.5$) is possible. With dark energy, a transient acceleration is mathematically impossible and an early loitering period seems to agree with the data if $\Gamma=O(10^{-1})$. Note that the DGP theory with minus sign and only CDM, disagrees with the data if one adds dark energy. Then a transient acceleration is still mathematically impossible and an early loitering period agrees with the data for positive values of $\Gamma$.\\
\begin{table*}
\begin{center}
\caption[]{Values corresponding to the smallest $\chi^2$ for each brane theories in agreement with and reasonably constrained by both the supernovae and CMB data.}
\begin{tabular}{lllllll}
\hline
Model & prior & $\chi^2_{DOF}$ & $\Omega_{m_0}$ & $\Omega_{k_0}$ & $\Omega_{e_0}$ & Others\\
\hline
Card. 1 & $\Omega_{m_0}=0.27\pm0.05$ & 1.16 & $0.29$ & - & - & $\nu=-1.01$\\
      2 & $\Omega_{m_0}=0.27\pm0.05$ & 1.17 & $0.30$ & $-0.02$ & - & $\nu=-0.98$\\
	    & and $\Omega_{k_0}=-0.02\pm0.05$ &  &  &  &  & \\
\hline
\hline
RS      & $\Omega_{d_0}=0\pm0.1$ &  & $0.15$ & - & $0.8$ & $\Omega_{d_0}=-0.008$ \\
\hline
\hline
DGP 1   & $+$ sign, $\Omega_{m_0}=0.27\pm0.05$ & $1.17$ & $0.23$ & $-0.04$ & - & -\\
        & and $\Omega_{k_0}=-0.02\pm0.05$ &  &  &  &  & \\
    2   & $+$ sign, $\Omega_{m_0}=0.27\pm0.05$ & $1.14$ & $0.27$ & - & $0.17$ & $\Gamma=-12.83$\\
\hline
\end{tabular}
\label{tabChi2A}
\end{center}
\end{table*}

\begin{table*}
\begin{center}
\caption[]{Values corresponding to the smallest $\chi^2$ for each brane theories in agreement with and reasonably constrained by both the supernovae and CMB data.}
\begin{tabular}{llllllll}
\hline
Model & $H_0$ & Age & $z_a$ & $z_d$ & Trans. & Loit.\\
\hline
Card. 1 & $64.6$ & $14.64$ & $0.69$ & $0.34$ & No & Yes($z_l<1$) \\
      2 & $64.6$ & $14.63$ & $0.69$ & $0.32$ & No & Yes($z_l<1$) \\
\hline
\hline
RS     &  $64.7$ & $11.4$ & $0.44$ & $0.34<z<1.17$ & No & Yes (early-time) \\
	   &   		 & 		  & 	   & 	           &    & if $\Omega_{\Lambda_0}<0.8$ \\
\hline
\hline
DGP 1  & $64.86$ & $14.96$ & $0.76$ & $0.68$ & No & Yes($z_l<1.5$)\\
    2  & $67.75$ & $14.13$ & $0.32$ & $0.33$ & No & Yes(early-times)\\
	   &   		 & 		   & 	    & 	     &    & if $\Gamma=O(10^{-1})$\\
\hline
\end{tabular}

\label{tabChi2B}
\end{center}
\end{table*}
Hence, a transient acceleration and an early loitering period - two interesting properties of branes - usually disagree with the data. In particular, the loitering period usually occurs at late time. All the theories shown in Tab. \ref{tabChi2A}-\ref{tabChi2B} predict a Hubble constant around $65$ (around $68$ for the DGP model with curvature) and an age for the universe around $15$ billions years (around $11.4$ for the Randall-Sundrum model).\\ Moreover, all the brane models we have considered are equivalent to GR with dark energy defined by a barotropic index $\gamma_\phi$. Each time one takes the supernovae and CMB data into account, this dark energy is close to the form of a cosmological constant, i.e. $\gamma_\phi=0$. Hence for the Cardassian model, $\gamma_\phi\approx 10^{-2}$. For the Randall-Sundrum and DGP models, their barotropic index varies but is also close to $0$ at present time. Indeed, their barotropic index tends to some constants in the past and in the future, these constants being related by a fast transition period and the constant value $\gamma_\phi=0$ reached in the future (past) for the Randall-Sundrum (DGP) model. What is striking is that the data predict we should live just during the short transition period. What does it mean? Probably that the value of $\gamma_\phi$ is so close to $0$ today that we are not able to measure it precisely enough to prove that we are in the asymptotical period when $\gamma_\phi\rightarrow 0$ instead of the transition period where $\gamma_\phi$ may be slightly different from zero (in the same way that if the universe is nearby flat, we can prove it only if the measures are precise enough to measure a very very sligth curvature). This is probably why the data lead us to believe that we are living during a transition period when $\gamma_\phi$ is close to zero and not yet in an asymptotical epoch when $\gamma_\phi\rightarrow 0$.\\\\
To conclude, if our universe is described by one of the three brane-inspired models studied in this paper, today it is probably very similar to a $\Lambda CDM$ universe. But how similar? The supernovae data are not precise enough to answer (and consequently could fake a universe undergoing a transition to a $\Lambda CDM$ model instead of being already asymptotically enter into such a state) but it could be the case of structures formation data (as SDSS or 2dFGRS) as recently shown in \cite{AmaElgMul04}.
\appendix
\section{Minimalisation, marginalisation and priors}\label{l1}
The relation between the apparent magnitude, the magnitude zero point offset and the Hubble free luminosity distance writes
$$
m=\bar M+5log(D_l)
$$
with
$$
D_l=(1+z)\int1/\sqrt{E(z)}dz
$$
and 
$$
\bar M=M+5log(\frac{c}{H_0})+25
$$
One can define the $\chi^2$ as
$$
\chi^2=\sum_{p=1}^{nb}\frac{(m^{obs}_i-m^{th}_i)^2}{\sigma^2_i}
$$
where $nb$ is the number of data. Let us consider a cosmological model defined by $n$ parameters. The model fits the data best when the values of its $n$ parameters and the value of the constant $\bar M$ minimise $\chi^ 2$.\\
Another way to find the best fit consists in marginalising a nuisance parameter, say $\bar M$, since in the Riess data, the Hubble constant $H_0$ is unknown. To marginalise, one defines a new $\bar \chi^2$ as
$$
\bar \chi^2=-2\ln\int^{+\infty}_{-\infty}e^{-\chi^2/2}d\bar M
$$
After some algebraic manipulation and after defining 
$$
A=\sum_{p=1}^{157}\frac{(m^{obs}_i-5log(D_l))^2}{\sigma^2_i}
$$
$$
B=\sum_{p=1}^{157}\frac{m^{obs}_i-5log(D_l)}{\sigma^2_i}
$$
$$
C=\sum_{p=1}^{157}\frac{1}{\sigma^2_i}
$$
it becomes 
\begin{equation}\label{chi22}
\bar \chi^2=A-\frac{B^2}{C}+ln\frac{C}{2\pi}.
\end{equation}
Expanding $\chi^2$ one also easily finds that 
$$
\chi^2=A-2\bar MB+\bar M^2C.
$$
Since minimalisation corresponds to $d\chi^2/d\bar M=0$, it also occurs for $\bar M=\frac{B}{C}$, that is 
\begin{equation}\label{chi21}
\chi^2=A-\frac{B^2}{C}
\end{equation}
Both (\ref{chi21}) and (\ref{chi22}) define two $n$ surfaces containing the values $(i_1,i_2,...,i_n)$ of the $n$ cosmological model parameters minimising $\chi^2$ and $(j_1,j_2,...,j_n)$ of the $n$ cosmological model parameters marginalising $\bar\chi^2$.\\
The first $n$ values are solutions of 
\begin{equation}\label{as1}
\frac{d\chi_2}{di_k}=0.
\end{equation}
The second $n$ values are solutions of
\begin{equation}\label{as2}
\frac{d\bar\chi_2}{dj_k}=0.
\end{equation}
Since $\chi_2$ and $\bar\chi_2$ only differ by a constant, these two groups of equations have the same solutions and $(i_1,i_2,...,i_n)=(j_1,j_2,...,j_n)$.\\
Moreover, we have $\bar \chi^2_{min}-\chi^2_{min}= \ln\frac{C}{2\pi}$. The more accurate the observations, and the smaller the $\sigma_i$, the larger will be the difference between $\chi^2_{min}$ and $\bar\chi^2_{min}$. Thus evaluating the confidence contours using minimalisation or marginalisation is equivalent, as long as (\ref{as1}-\ref{as2}) hold.\\
Sometimes, the confidence contours are too large or the parameter values giving the best (i.e. the smallest) $\chi^2$ are unphysical. In this case, it may be interesting to assume some priors on one or several parameters of the model. For instance, one can assume that the $CDM$ density parameter $\Omega_{m_0}$ should be $\Omega_{m_0}=0.27\pm0.05$. One way to assume this prior mathematically consists in minimalising the quantity $\chi^2_{prior}=\chi^2+(\Omega_{m_0}-0.27)^2/0.05^2$. The same method applies for marginalising with a prior.
\section*{Acknowledgements}
I thank E. Di Pietro and D. Pelat for useful discussions of the $\chi^2$ test and V. Sahni for interesting discussions on loitering. This work is supported by the Marie Curie Intra-European Fellowship Programs of the European Union Commission (MEIF-CT-2005-515028).

\bibliographystyle{aa}

\end{document}